\documentclass[a4paper,11pt]{article}
\pdfoutput=1
\usepackage{jheppub}
\usepackage[T1]{fontenc}
\usepackage{amsmath,amssymb,amsthm,mathtools,booktabs,array,longtable}

\newcommand{\h}{\hbar}
\newcommand{\dd}{\mathrm d}
\newcommand{\Res}{\operatorname*{Res}}
\newcommand{\res}{\operatorname{res}}
\newcommand{\tr}{\operatorname{tr}}
\newcommand{\Rhat}{\widehat R}
\newcommand{\Qhat}{\widehat Q}
\newcommand{\cD}{\mathcal D}
\newcommand{\cL}{\mathcal L}
\newcommand{\cB}{\mathcal B}
\newcommand{\cF}{\mathcal F}
\newcommand{\vecc}{\mathrm{vec}}
\newtheorem{theorem}{Theorem}[section]
\newtheorem{proposition}[theorem]{Proposition}
\newtheorem{conjecture}[theorem]{Conjecture}
\numberwithin{equation}{section}
\allowdisplaybreaks[2]
\title{Resolvent reconstruction of minimal strings beyond KdV II: D series models}

\author[a]{Wasif Ahmed,}
\author[b]{Joydeep Naskar}
\affiliation[a]{Physics Department, Broida Hall, University of California,\\ Santa Barbara, CA 93106, USA}
\affiliation[b]{Beijing Institute of Mathematical Sciences and Applications,\\ Huaibei Town, Huairou District, Beijing 101408, China}
\emailAdd{wasif@ucsb.edu}
\emailAdd{joydeepnaskar@bimsa.cn}

\abstract{Three-state Potts gravity and tetracritical Ising gravity have the same central charge and the same conformal spectral curve. What distinguishes their amplitudes once a handle or a boundary is added? We answer this question by extending Johnson's differential-equation approach, in the reconstruction form developed in Part I \cite{AhmedNaskar2026} for the reduced KP hierarchies, to D-series minimal strings. Their spatial operator is a pseudo-differential pencil with a nonlocal Pfaffian term, and the classical energy acquires a second pole at zero momentum. Regularity at both poles fixes the quantum corrections uniquely, ordinary string equations supply them, and the ordinary D flows add boundaries in a way that matches the tau function. At genus one the closed free energy follows from the pencil as a sum of logarithms of critical-value velocities. The two theories differ through the origin of the curve, which the D theory counts as two canonical directions, and through the pairing of every point with its reflection. We follow the origin contribution through the free energies to genus three, the one-holed torus and the torus one-point amplitudes, and in the primary range the latter reproduce the formulas conjectured from worldsheet numerics for both series.}


\begin{document}
\maketitle
\flushbottom

\section{Introduction}\label{sec:intro}

A surface can change in two simple ways. It can gain a handle, or it can gain another boundary. In a double-scaled matrix model both operations are encoded in a single function $u(x)$, the solution of a string equation \cite{BrezinKazakov1990,DouglasShenker1990,GrossMigdal1990,Douglas1990,DiFrancescoGinspargZinnJustin1995}. Its expansion in the renormalized coupling $\h$ adds handles, while a macroscopic loop, which opens a boundary, is computed from the resolvent of the Schr\"odinger operator $-\h^2\partial_x^2+u(x)$ \cite{BanksDouglasSeibergShenker1990,Moore:1991ir}. In this paper we ask how far that old picture extends.

Two years ago Johnson observed that the picture gives an efficient route to one-boundary amplitudes \cite{Johnson2024Virasoro,Johnson2024SuperVirasoro}. One expands the diagonal resolvent $\Rhat(x,E)$ at energy $E$ in powers of $\h$, using the Gel'fand--Dikii equation \cite{GelfandDikii1975}, and substitutes the genus expansion of $u(x)$ dictated by the string equation. At every positive genus the coefficient of $\h^{2g}$ turns out to be a total derivative,
\begin{equation}\label{eq:intro-totalderivative}
 \Rhat_g(x,E)=\partial_x\Qhat_g(x,E).
\end{equation}
The integral over the Fermi sea that defines the boundary amplitude $W_{g,1}$ then collapses to an evaluation of $\Qhat_g$ at the Fermi level $x=\mu$. Since $\Qhat_g$ is built from the leading potential $u_0$ and finitely many of its derivatives, a one-boundary amplitude at any genus is fixed by a handful of numbers at a single point.

Ahmed, Johnson and Saraswat tested this ``ODE method'' extensively \cite{AhmedJohnsonSaraswat2025}. They pushed the genus expansion to high order, developed a descending construction of the primitives $\Qhat_g$, and used it to compute Weil--Petersson volumes for JT supergravity with extended supersymmetry. Their construction already contains a hint of what we exploit below. When the poles of the primitive are matched from the strongest to the weakest, the final equation determines the next correction to $u(x)$, so the primitive and the potential are fixed together. Johnson and Rodrigues have since used the same equations to obtain nonperturbative data for Weil--Petersson volumes and intersection numbers \cite{JohnsonRodrigues2026}.

As described so far, however, the method computes one boundary at a time. Macroscopic loops had been organized by the KdV flows before \cite{OkuyamaSakai2020,Lowenstein2024OpenClosed,Lowenstein2024Volumes}, and Johnson then showed how to add boundaries within the ODE method \cite{Johnson2025Kernel,Johnson2026Universal}. His loop operator $\delta_E$ is a generating function of KdV time derivatives. Acting on the closed free energy $F_g$ it produces $\Qhat_g$, and acting again it produces amplitudes with more boundaries. On functions of $u_0$ and its derivatives the operator becomes elementary, so the amplitudes $W_{g,k}$ with $k$ boundaries become universal formulae in $u_0(\mu)$ and its derivatives. These formulae cover the Airy and Bessel models, Weil--Petersson volumes \cite{Mirzakhani2007,EynardOrantin2007WP,SaadShenkerStanford2019} and their supersymmetric cousins \cite{Stanford:2019vob,Turiaci:2023jfa}, and the $(2,2l-1)$ minimal strings, all of which are governed by the KdV hierarchy, the two-reduction of KP.

Does the method survive when the auxiliary operator is not of Schr\"odinger type? In Part I of this series \cite{AhmedNaskar2026} we answered this question for the $r$-reduction of KP, the $A_{r-1}$ Drinfeld--Sokolov hierarchy, which contains the $(p,p')$ minimal strings with $p>2$ and in particular the Ising model on random surfaces \cite{Kazakov1986}. There we found that the method can be turned around. Instead of starting from a solution of a string equation, we start from the classical spectral curve together with its motion in $x$, and ask that the integrated resolvent behave like a boundary amplitude. Its poles should sit only where two sheets meet, and the amplitude should be regular at infinity. These two conditions determine the quantum corrections to every field of the operator, genus by genus, and a string equation guarantees that a solution exists. The idea is close in spirit to the reconstruction of quantum curves from topological recursion \cite{Bouchard:2016obz,Eynard:2021sxg}, with the ordering and the genus expansion of the operator fixed in advance. Moreover, the same resolvent leads to amplitudes with several boundaries \cite{BergereEynard2009,BergereBorotEynard2015,BelliardEynardMarchal2016,BelliardEynardMarchal2018}.

In this paper we take the next step and turn to the D series. Minimal models \cite{BelavinPolyakovZamolodchikov1984} come with an ADE classification of their modular invariants \cite{CappelliItzyksonZuber1987}. The A series contains the diagonal theories, while the D series contains theories such as the three-state Potts model. Potts has the same central charge as the tetracritical Ising model, $c=4/5$, but a different spectrum and a different torus partition function. Coupled to Liouville gravity \cite{Knizhnik:1988ak,David:1988hj,Distler:1988jt}, D-series models give D-series minimal strings, whose matrix-model duals are expected to involve several matrices and have not been solved. Rodriguez, Usatyuk and Wang have recently computed many of their sphere and torus amplitudes by direct integration over the moduli space and conjectured closed formulas for the torus one-point amplitudes \cite{RodriguezUsatyukWang2025}, which gives us a sharp target.

The integrable system behind the D series is the D-type Drinfeld--Sokolov hierarchy \cite{DrinfeldSokolov1985,LiuWuZhang2009}, whose spatial operator is no longer differential but a pseudo-differential pencil,
\begin{equation}\label{eq:intro-pencil}
 L_D=D^{h+1}+\sum_{i=1}^{n-1}\frac12\bigl\{u_i,D^{2i-1}\bigr\}+\rho D^{-1}\rho,
 \qquad L_D\psi=\lambda D\psi,\qquad h=2n-2,
\end{equation}
with $D=\h\partial_x$ and spectral parameter $\lambda$. The integer $n$ labels the Lie algebra $D_n$, whose Coxeter number is $h$, and an additional field $\rho$ enters through a nonlocal term. Every field has an even expansion in $\h$, and we write $u_{i,0}$ and $\rho_0$ for the leading terms. Two things change compared with Part I. First, the classical energy acquires a pole at zero momentum,
\begin{equation}\label{eq:intro-curve}
 W(p,x)=p^h+\sum_{i=1}^{n-1}u_{i,0}(x)\,p^{2i-2}+\frac{\rho_0(x)^2}{p^2},
\end{equation}
so the integrated resolvent must now be regular at two places. Second, the new field, which we call the Pfaffian field, vanishes at the conformal point of the minimal string, where the D curve becomes a polynomial in $p^2$ and coincides with an A curve. This raises a puzzle that will accompany us through the paper. If Potts and tetracritical Ising share the same classical curve at the conformal point, where does the difference between them come from?

The answer lies in how the D theory treats the origin and the reflection $p\to-p$. When $\rho_0$ is small, the critical point of $W$ at $p=0$ resolves into four nearby momentum points, which form two canonical directions whose critical values differ by an amount of order $\rho_0$ and coincide only in the limit. On the A side the same point counts once. In the limit $\rho_0\to0$, however, the D reconstruction remembers both directions, and this is enough to separate the two theories at every order we have computed, through genus three. At genus one, for instance, the free energy is a sum of logarithms of critical-value velocities, one for each canonical direction. The origin is therefore counted twice in D and once in A, and the logarithmic coefficients of the two theories end up in the ratio of their numbers of canonical directions. The origin enters the bulk amplitudes that we compare with the worldsheet in the same way. The reflection, on the other hand, is most visible in the boundary amplitudes, where the D cylinder pairs every point with its image.

Let us summarize what we find.
\begin{itemize}
 \item \emph{The recipe.} Away from $\rho_0=0$, regularity at the two poles of the energy map determines the quantum fields and the one-boundary primitive uniquely at every genus, and ordinary string equations supply a solution, as we sketch. Along normalized string families the fields continue analytically to the symmetric sector $\rho=0$.
 \item \emph{Boundaries.} The ordinary odd flows of the D hierarchy, collected into a loop operator, add boundaries. Their action on the tau function reproduces the reconstructed primitive at every genus. At genus zero they produce a cylinder paired with its image and the three-boundary residue formula of a mirror recursion. At higher order we sketch why the amplitudes have the pole structure and the linear loop equations of a topological recursion, leaving one quadratic loop equation open.
 \item \emph{Closed surfaces.} The genus-one free energy follows from the pencil itself, and the same three steps give the A-series result, up to a scalar tau-matching step that we do not write out. Exact data through genus three suggest formulas built from log-Hessians and their variations.
 \item \emph{Potts and tetracritical Ising.} We reconstruct both theories on a common family through genus three and isolate the origin contribution in the free energies, the one-holed torus and the torus one-point amplitudes.
 \item \emph{Worldsheet data.} In the primary range, our genus-one free energy reproduces the conjectured torus one-point formulas of ref.~\cite{RodriguezUsatyukWang2025} for both series, and in the same range the Pfaffian coupling reproduces their mixed sphere four-point amplitudes. Conditional on an identity for the Pfaffian Hessian, we also predict pure Pfaffian torus two-point amplitudes.
\end{itemize}

The paper is organized as follows. Section~\ref{sec:recipe} develops the one-boundary recipe, and section~\ref{sec:boundaries} explains how to add boundaries. Section~\ref{sec:closed} turns to closed surfaces. In section~\ref{sec:potts} we put everything to work on Potts and tetracritical Ising. We close with a summary and open questions in section~\ref{sec:discussion}. The appendices contain the proof sketches, the complete genus-one derivation, a regular low-rank example, the complete Potts data, further homogeneous backgrounds and the comparison with the worldsheet data of ref.~\cite{RodriguezUsatyukWang2025} in the primary range.

Table~\ref{tab:notation} collects the main symbols and where they are introduced.

\begin{table}[t]
\centering
\small
\renewcommand{\arraystretch}{1.12}
\begin{tabular}{@{}l>{\raggedright\arraybackslash}p{0.52\linewidth}>{\raggedright\arraybackslash}p{0.2\linewidth}@{}}
\toprule
Symbol & Meaning & Introduced\\
\midrule
$\h$, $g$, $k$ & genus-counting parameter, genus, number of boundaries & section~\ref{sec:intro}\\
$D$, $L_D$, $\cL$ & $D=\h\partial_x$, the pencil \eqref{eq:intro-pencil}, the Lax operator \eqref{eq:Lax} & sections~\ref{sec:intro} and~\ref{sec:curve}\\
$h$, $n$ & Coxeter number and rank of $D_n$, with $h=2n-2$ & section~\ref{sec:intro}\\
$u_i$, $\rho$ & ordinary fields and the Pfaffian field & section~\ref{sec:intro}\\
$x$, $\mu$ & background coordinate and the endpoint where amplitudes are evaluated & sections~\ref{sec:intro} and~\ref{sec:density}\\
$\lambda$, $p$ & spectral parameter, or energy, and momentum & sections~\ref{sec:intro} and~\ref{sec:curve}\\
$W(p,x)$ & classical energy \eqref{eq:intro-curve} & section~\ref{sec:intro}\\
$T$, $a$ & $T=p^3W_p$ and its roots, the ramification points & section~\ref{sec:curve}\\
$v_a$ & velocity of the critical value at $a$ & section~\ref{sec:curve}\\
$\cD$ & derivative in $x$ at fixed energy & section~\ref{sec:density}\\
$\Rhat_{2g}$, $\Qhat_g$ & sheet density at order $\h^{2g}$ and its primitive & section~\ref{sec:density}\\
$\Phi_g$ & variation of the energy by the new fields at genus $g$ & section~\ref{sec:density}\\
$\omega_{g,k}$ & amplitude of genus $g$ with $k$ boundaries & sections~\ref{sec:density} and~\ref{sec:genuszero}\\
$S$, $\pi$, $\kappa$, $y$ & string generator, its leading symbol, string constant, disk & section~\ref{sec:strings}\\
$\gamma_0$ & origin constant, the coupling of the Pfaffian field & section~\ref{sec:strings}\\
$\cB(\lambda)$ & loop operator that inserts a boundary at energy $\lambda$ & section~\ref{sec:loop}\\
$B_D$ & paired cylinder & section~\ref{sec:genuszero}\\
$F^D_g$, $F_*$, $v_{\rm fix}$ & free energy, velocity sum \eqref{eq:Fone}, velocity of the point $p=0$ & sections~\ref{sec:loop} and~\ref{sec:genusone}\\
$(h,s)$, $(r,s')$, $m$ & model labels, matter labels and the index $m=|sr-hs'|$ & sections~\ref{sec:strings} and~\ref{sec:pottsboundaries}\\
\bottomrule
\end{tabular}
\caption{Main symbols. Subscripts on $W$ denote partial derivatives, and primes on fields denote $x$-derivatives.}\label{tab:notation}
\end{table}
\section{The recipe for one boundary}\label{sec:recipe}

We begin with a single boundary, and the plan follows Part I closely. We first describe the classical curve that the D pencil produces and how it moves with the background coordinate $x$. We then build the resolvent, ask what analytic properties its integral must have to be a boundary amplitude, and show that these properties fix all quantum corrections. Finally we explain why string equations guarantee that the reconstruction never gets stuck, and what happens when the Pfaffian field is switched off.

\subsection{A moving curve with two poles}\label{sec:curve}

The operator \eqref{eq:intro-pencil} acts on functions of $x$, with $D=\h\partial_x$ and symmetrically ordered coefficients. The fields $u_1,\ldots,u_{n-1}$ multiply odd powers of $D$, and the Pfaffian field $\rho$ appears sandwiched around the formal inverse $D^{-1}$. The spectral parameter multiplies a derivative, $L_D\psi=\lambda D\psi$. Equivalently, the hierarchy operator
\begin{equation}\label{eq:Lax}
 \cL=D^{-1}L_D=D^{h}+\cdots+D^{-1}\rho D^{-1}\rho
\end{equation}
has eigenvalue $\lambda$. This is the Lax operator of the D-type Drinfeld--Sokolov hierarchy in the form used by Liu, Wu and Zhang \cite{LiuWuZhang2009}. Every field has an even expansion, $u_i=\sum_{g\geq0}\h^{2g}u_{i,g}$ and $\rho=\sum_{g\geq0}\h^{2g}\rho_g$.

At leading order we replace $D$ by a momentum $p$, and the pencil equation becomes $p\,(W(p,x)-\lambda)=0$, with the energy $W$ of \eqref{eq:intro-curve}. Two features distinguish it from the polynomial energies of Part I. It is even in $p$, so its sheets come in pairs $\pm p$, and it has a second pole, at $p=0$, whose strength is set by the Pfaffian field. For a generic energy there are $h+2=2n$ sheets. Subscripts on $W$ denote partial derivatives, and two sheets meet at a zero of $W_p$. These ramification points are the roots of the polynomial
\begin{equation}\label{eq:T}
 T(p)=p^3W_p(p),
\end{equation}
which has degree $2n$. The roots come in opposite pairs $\pm a$, and each pair is one canonical direction of the curve, so there are $n$ canonical directions. For comparison, the scalar operator of Part I with $h$ sheets has $h-1$ ramification points, and this difference in counting is what will separate Potts from tetracritical Ising in section~\ref{sec:potts}.

The curve is not static, since the fields depend on $x$ and so do the positions of the ramification points and their energies. A critical value $W(a(x),x)$ moves with velocity
\begin{equation}\label{eq:velocity}
 v_a=\frac{\dd}{\dd x}W(a(x),x)=W_x(a),
\end{equation}
since $W_p(a)=0$. We work on the regular domain
\begin{equation}\label{eq:regular}
 \rho_0\neq0,\qquad T\ \text{has simple roots},\qquad v_a\neq0\ \text{at every root}.
\end{equation}
The first condition keeps the pole at the origin, the second makes every ramification simple, and the third says that every critical value moves.

\subsection{The resolvent and its integral}\label{sec:density}

The quantum problem is to invert $L_D-\lambda D$. To keep track of operator ordering we work with Weyl symbols, for which composition becomes the Moyal star product
\begin{equation}\label{eq:star}
 f_1\star f_2=f_1\exp\!\left[\frac\h2\bigl(\overleftarrow\partial_p\overrightarrow\partial_x-\overleftarrow\partial_x\overrightarrow\partial_p\bigr)\right]f_2 .
\end{equation}
The symbol of $L_D-\lambda D$ is $\mathsf A_{\rm W}=p(W-\lambda)+\h^2\mathsf A_2+\cdots$, where the correction $\mathsf A_2$ contains two kinds of terms. One is the first quantum correction of the fields themselves, and the other comes from ordering, because commuting $D$ past a field produces derivatives. For the nonlocal term the ordering correction is
\begin{equation}\label{eq:rho-symbol}
 \rho\star p^{-1}\star\rho=\frac{\rho^2}{p}+\frac{\h^2}{2p^3}\bigl(\rho\rho''-(\rho')^2\bigr)+O(\h^4),
\end{equation}
with primes denoting $x$-derivatives. Small as it is, this term is what makes the genus-one construction of section~\ref{sec:closed} close at the origin.

The inverse symbol $G$, defined by $\mathsf A_{\rm W}\star G=G\star \mathsf A_{\rm W}=1$, is found order by order. Its leading term is $1/\mathsf A_0$ with $\mathsf A_0=p(W-\lambda)$, and the first correction is
\begin{equation}\label{eq:G2}
 G_2=-\frac{\mathsf A_2}{\mathsf A_0^2}+\frac{\mathsf A_{0,pp}\mathsf A_{0,xx}-\mathsf A_{0,px}^2}{4\mathsf A_0^3}
 -\frac{\mathsf A_{0,pp}\mathsf A_{0,x}^2-2\mathsf A_{0,px}\mathsf A_{0,p}\mathsf A_{0,x}+\mathsf A_{0,xx}\mathsf A_{0,p}^2}{4\mathsf A_0^4},
\end{equation}
with all derivatives at fixed $\lambda$. The diagonal resolvent on a given sheet is obtained by a residue in $p$ at the sheet root $p_\lambda$, the solution of $W(p_\lambda,x)=\lambda$ on that sheet. Because the spectral parameter multiplies $D$, the natural density carries one factor of the momentum,
\begin{equation}\label{eq:density}
 \Rhat=\Res_{p=p_\lambda}\bigl(pG\,\dd p\bigr),\qquad \Rhat_0=\frac1{W_p}.
\end{equation}
The residues are elementary, since a term with $(W-\lambda)^{-j}$ is replaced by $j-1$ applications of $\partial_\lambda=W_p^{-1}\partial_p$, divided by $(j-1)!$. The result $\Rhat=\sum_g\h^{2g}\Rhat_{2g}$ is a rational function of the leading momentum on the sheet, and its poles sit where $W_p=0$. Reversing $\h$ and reversing $p$ together show that each coefficient $\Rhat_{2g}$ is odd in $p$.

\label{sec:endpoints}
To turn the density into an amplitude we have to integrate it in $x$ at fixed energy. As $x$ changes, the momentum on a sheet must move to keep $W(p,x)=\lambda$, so the relevant derivative is
\begin{equation}\label{eq:fixedenergy}
 \cD=\partial_x\big|_\lambda=\partial_x\big|_p-\frac{W_x}{W_p}\partial_p,
 \qquad \cD\Qhat_g=\Rhat_{2g}.
\end{equation}
This is the analogue of the total-derivative relation \eqref{eq:intro-totalderivative}. The one-boundary amplitude is the primitive evaluated at the endpoint $x=\mu$ and dressed by the energy differential,
\begin{equation}\label{eq:omega}
 \omega_{g,1}=W_{g,1}(p)\,\dd p=\Qhat_g(p,x)\,W_p(p,x)\,\dd p\,\Big|_{x=\mu}.
\end{equation}

Which primitive should we take? A boundary amplitude of a minimal string is a meromorphic differential on the spectral curve whose only singularities sit at ramification points. The energy map has two poles, at $p=\infty$ and $p=0$, and the differential must be regular at both. Since $W_p\sim hp^{h-1}$ at infinity and $W_p\sim-2\rho_0^2p^{-3}$ at zero, this translates into
\begin{equation}\label{eq:endpoints}
 \Qhat_g=O\bigl(p^{-h-1}\bigr)\ \ (p\to\infty),\qquad
 \Qhat_g=O\bigl(p^{3}\bigr)\ \ (p\to0),
\end{equation}
together with the requirement that $\Qhat_g$ have poles only at the roots of $T$, of order at most $6g-1$. The first condition is the one used in Part I, while the second is new. It is the price of the nonlocal term and, as we will see, also what fixes the Pfaffian field.

Finally, at genus $g$ the fields acquire new corrections $u_{i,g}$ and $\rho_g$, which we collect into the variation of the leading energy,
\begin{equation}\label{eq:Phi}
 \Phi_g(p)=\sum_{i=1}^{n-1}u_{i,g}\,p^{2i-2}+d_g\,p^{-2},\qquad d_g=[\h^{2g}]\rho^2 .
\end{equation}
Varying $W$ at fixed energy displaces the sheet, and the density changes accordingly,
\begin{equation}\label{eq:Rvariation}
 \Rhat_{2g}=\mathcal S_g+\frac{\Phi_gW_{pp}-\Phi_g'W_p}{W_p^3},
\end{equation}
where $\Phi_g'=\partial_p\Phi_g$ and the source $\mathcal S_g$ is determined by lower genera. The reconstruction problem is now to find, given the moving curve and all lower-genus data, a field variation $\Phi_g$ and a primitive $\Qhat_g$ that satisfy \eqref{eq:fixedenergy} and the endpoint conditions \eqref{eq:endpoints}.

\subsection{Uniqueness of the solution}\label{sec:unique}

Why should such a problem have at most one solution? The argument is the one that already worked for KdV and for the reduced KP hierarchies, and it starts from the strongest poles. The new fields enter \eqref{eq:Rvariation} only with poles of order at most three at a ramification point, whereas $\cD$ raises the order of a pole by two, with a leading coefficient proportional to the velocity $v_a$. As long as the critical value moves, the most singular part of $\Qhat_g$ is therefore fixed by the source alone. Subtracting it and repeating, we descend through the pole orders until only the new fields and a simple-pole remainder are left.

Suppose now that two solutions exist, and look at the difference $\delta\Qhat$ of their primitives. It can only have simple poles at the roots of $T$, so $\delta\Qhat=N(p)/T(p)$ with a polynomial $N$. Decay at infinity forces $\deg N\leq1$, and regularity at the origin forces $p^3\mid N$, because $T(0)=-2\rho_0^2\neq0$. Hence $\delta\Qhat$ vanishes, and the remaining equation $\partial_p(\Phi/W_p)=0$ for the difference of field variations gives $\Phi=c(x)W_p$. This is impossible, since $\Phi$ is even and $W_p$ is odd, and so the reconstruction is unique.

To see what the condition at the origin does, let us drop it. A one-parameter ambiguity then reappears, proportional to
\begin{equation}\label{eq:pfkernel}
 \Bigl(\partial_\lambda\frac{\rho_0}{p},\ \Bigl\{\frac{\rho_0}{p},W\Bigr\}\Bigr),\qquad \{f_1,f_2\}=\partial_pf_1\,\partial_xf_2-\partial_xf_1\,\partial_pf_2,
\end{equation}
a simultaneous change of the primitive and of the field correction, which is the first dispersionless Pfaffian flow of the hierarchy. It satisfies the condition at infinity but carries a forbidden linear term at zero. The origin condition therefore fixes the integration constant of the Pfaffian sector, just as the condition at infinity fixes the ordinary ones.

On a homogeneous background the descent becomes finite linear algebra. Writing $\Qhat_g$ as an odd numerator over $(p^3W_p)^{6g-1}$, the two endpoint conditions cut the numerator down to $n(6g-2)$ coefficients. Together with the $n$ new fields this gives $n(6g-1)$ unknowns, against $6ng$ equations from \eqref{eq:fixedenergy}. Uniqueness tells us that the unknowns are independent, but it leaves $n$ compatibility conditions, and whether they hold is the question of existence.

\subsection{String equations}\label{sec:strings}

Existence comes from the hierarchy, which has two families of commuting flows \cite{LiuWuZhang2009}. The ordinary flows are generated by $B_j=(P^j)_+$, where $P=\cL^{1/h}$ is the root that begins with $D$ and $j$ is odd, while the Pfaffian flows are generated by powers of a second root $\widehat P$ that begins with $D^{-1}\rho$. An ordinary string background is a finite combination
\begin{equation}\label{eq:string}
 S=\sum_{j\ {\rm odd}}\mathsf T_jB_j,\qquad [S,\cL]=\kappa\h,
\end{equation}
with constant coefficients $\mathsf T_j$ and a constant $\kappa\neq0$. At leading order the commutator becomes a Poisson bracket,
\begin{equation}\label{eq:pi}
 \pi=\sum_{j\ {\rm odd}}\mathsf T_j\bigl(W^{j/h}\bigr)_+,\qquad \{\pi,W\}=\kappa,
\end{equation}
where $+$ keeps the polynomial part at $p=\infty$. The partner $\pi$ also gives the disk, $\omega_{0,1}=y\,\dd W$ with $y=-\pi/\kappa$. A single-flow background takes $S=B_s$, and for a minimal string the partner degree $s$ is odd and coprime to $h$.

Something new happens in the Pfaffian sector. The ordinary generators obey $S^*=-DSD^{-1}$ and $S(1)=0$, so $\mathcal A=SD^{-1}$ is a self-adjoint differential operator, and the Pfaffian component of the string equation integrates once to
\begin{equation}\label{eq:origin}
 \mathcal A(\rho)=\gamma_0(\h),\qquad \rho_0\,\pi_p(0)=\gamma_0^{(0)} .
\end{equation}
The constant $\gamma_0$ is an additional integration constant, so an ordinary string can carry a nonzero Pfaffian field, and $\gamma_0$ measures how far the background sits from the symmetric sector.

A solution of the string equation provides a primitive automatically. Localizing the pencil with one auxiliary component $\chi=D^{-1}(\rho\psi)$ turns it into a first-order system of size $2n$, and the string equation supplies a compatible connection in $\lambda$. The trace of that connection against the sheet projector integrates the density exactly, in close analogy with the trace primitive of Part I, once it is corrected by a term that comes from ordering the momentum factor of the density. We call this term the \emph{ordering subtraction}. Appendix~\ref{app:reconstruction} sketches why the positive-genus coefficients of this corrected trace satisfy the growth conditions, apart from finitely many terms that are constant in $x$. These are removed by the integration constants of the string, $n-1$ at infinity and one at the origin. A normalized string solution is therefore an admissible solution of the reconstruction problem, and uniqueness identifies the two.

The last step promotes this to a statement about arbitrary moving curves. With enough ordinary flow coefficients, the leading jets of string solutions fill an open set, and every compatibility residual is a rational function of finitely many jets that vanishes on this open set, and therefore vanishes identically.

\begin{proposition}[Regular reconstruction]\label{thm:regular}
For an analytic moving D curve satisfying \eqref{eq:regular}, the reconstruction problem has a solution at every genus. It is unique, and its fields and primitive are rational functions of finitely many jets of the leading fields.
\end{proposition}
Uniqueness is proved in section~\ref{sec:unique}, and for existence we give a sketch in appendix~\ref{app:reconstruction}, which states the endpoint estimates and shows the induction at the origin and the open-jet argument.

\subsection{Turning off the Pfaffian field}\label{sec:symmetric}

The minimal strings we care about sit at $\rho=0$. This symmetric sector is invariant under the ordinary flows, and there the curve becomes a polynomial in $p^2$ whose origin is an ordinary critical point. The D curve at $\rho=0$ therefore looks like an A curve with even energy, and the regular domain \eqref{eq:regular} has been left behind. What does the reconstruction do in this limit?

We can see the answer through a two-variable potential,
\begin{equation}\label{eq:LG}
 \cF(t,\mathsf y)=A(t)-t\mathsf y^2+2\rho\mathsf y,\qquad A(t)=t^{h/2}+\sum_{i=1}^{n-1}u_it^{i-1}.
\end{equation}
For $\rho\neq0$, eliminating $\mathsf y$ through $t\mathsf y=\rho$ and setting $t=p^2$ returns the energy $W$. For $\rho=0$ the elimination is singular, but $\cF$ itself is regular, with critical points at the $n-2$ zeros of $A'(t)$ at $\mathsf y=0$, together with the two Morse points $(t,\mathsf y)=(0,\pm\sqrt{A'(0)})$, which share a critical value. In momentum space the same thing happens through a collision. If we write the symmetric energy near the origin as $W_0+W_2p^2+O(p^4)$, adding the Laurent term gives four momentum points with
\begin{equation}\label{eq:cluster}
 p^4=\frac{\rho_0^2}{W_2}+O(\rho_0^3),\qquad
 W(p)=W_0\pm2\sqrt{W_2}\,\rho_0+O(\rho_0^2),\qquad W_{pp}(p)\to8W_2 .
\end{equation}
The four points form two opposite pairs, so they carry two canonical directions whose critical values coincide in the limit, and we call this configuration the \emph{resolved origin}.

The question has now become one of analytic continuation. Take a family of normalized ordinary strings whose origin constant $\gamma_0=\eta$ is a small parameter and whose ordinary constants are analytic and even in $\eta$. On the regular side the reconstruction is unique. As $\eta\to0$, do the fields have a limit, and is that limit the reconstruction in the symmetric sector?

\begin{proposition}[Symmetric limit]\label{thm:symmetriclimit}
Let the ordinary constants of the family be analytic\footnote{Analyticity is needed, since an even smooth deformation such as $e^{-1/\eta^2}$ has vanishing derivatives at $\eta=0$.} and even in $\eta$. Suppose that at $\eta=0$ the full classical hodograph Jacobian of the string, the Jacobian of its $n$ leading integrated equations with respect to the $n$ leading fields, is invertible, $\pi_p(0)\neq0$, and $W_p$ has simple zeros with nonzero velocities, including the zero at $p=0$. Normalize every positive-genus endpoint constant to zero. Then the regular reconstruction extends analytically in $(x,\eta)$ at each genus, with $u_{i,g}$ even and $\rho_g$ odd in $\eta$. Its limit is the unique solution in the sector $\rho=0$ whose primitive may have poles at the limiting critical points, including $p=0$.
\end{proposition}

The proposition rests on the structure of the equations at each genus. There the new fields solve a linear system whose matrix is the classical hodograph Jacobian and whose right side is a differential polynomial with coefficients regular at $\rho=0$, so analyticity of the inverse Jacobian propagates from one genus to the next. Appendix~\ref{app:symmetriclimit} sketches the argument and checks the hypotheses at the Chebyshev points that describe the conformal minimal strings.

The proposition tells us that the symmetric sector is reached as a limit rather than entered from scratch, and this is the key to the comparison with the A series. The A theory with the same even energy treats $p=0$ as a single critical point, whereas the D theory, reached from $\rho_0\neq0$, remembers that this point came from two canonical directions. Sections~\ref{sec:closed} and~\ref{sec:potts} follow that memory into the free energies and amplitudes, but first we need a way to add boundaries.
\section{Adding boundaries}\label{sec:boundaries}

So far the primitive has been a function of the fields along one background. To add a boundary by differentiation, however, we need to know how the primitive changes when the background itself is deformed by the hierarchy flows. Johnson's insight for KdV is that a boundary insertion is a particular combination of time derivatives, and that on the closed free energy it acts as the one-boundary primitive \cite{Johnson2026Universal}. We follow the same route for the D hierarchy, where the questions are which flows to combine, with which coefficients, and whether the result agrees with the reconstruction of section~\ref{sec:recipe} at every genus.

\subsection{Boundary insertion from the flows}\label{sec:loop}

Extend the fields to a local solution of all the commuting D flows, $\boldsymbol u(x,\boldsymbol t,\widehat{\boldsymbol t};\h)$, with ordinary times $t_j$, Pfaffian times $\widehat t_k$ and $t_1=x$, and write the logarithm of the tau function as $\log\tau_D=\sum_{g\geq0}\h^{2g-2}F^D_g$. In the normalization of Liu, Wu and Zhang \cite{LiuWuZhang2009}, its mixed second derivatives are residues of powers of the two roots, where $\res_D$ extracts the coefficient of $D^{-1}$,
\begin{equation}\label{eq:tau}
 2\h^2\partial_x\partial_{t_j}\log\tau_D=\res_DP^j,\qquad
 2\h^2\partial_x\partial_{\widehat t_k}\log\tau_D=\res_D\widehat P^{\,k}.
\end{equation}
In particular the top field is a second derivative of the free energy,
\begin{equation}\label{eq:taunorm}
 u_{n-1,g}=2h\,\partial_x^2F^D_g .
\end{equation}

An insertion is a linear combination of time derivatives with energy-dependent coefficients, which we find by comparing its $x$-derivative with the density \eqref{eq:density}. At large energy put $\mathsf z=\lambda^{1/h}$ and invert the energy map on the ordinary sheet, $p(\mathsf z)=\mathsf z+\sum_{j\ {\rm odd}}\varsigma_j\mathsf z^{-j}$. The classical density then expands as
\begin{equation}\label{eq:classicaldensity}
 \frac1{W_p}=\partial_\lambda p(\mathsf z)=\frac1h\mathsf z^{1-h}-\frac1h\sum_{j>0\ {\rm odd}}j\varsigma_j\,\mathsf z^{-h-j},
\end{equation}
where $-j\varsigma_j$ is the leading value of $\res_DP^j$. Comparing with \eqref{eq:tau} singles out
\begin{equation}\label{eq:loopoperator}
 \cB(\lambda)=\frac2h\sum_{j>0\ {\rm odd}}\lambda^{-1-j/h}\,\partial_{t_j}.
\end{equation}
Only the ordinary odd flows appear, and the Pfaffian flows are not needed for this boundary. This is natural, because the expansion at large energy only sees the ordinary sheet. The leftover term $\mathsf z^{1-h}/h$ in \eqref{eq:classicaldensity} is independent of the fields and will reappear in the cylinder.

On functions of the leading fields the insertion acts through the dispersionless flows. Summing the odd part of the Faber expansion,
\begin{equation}\label{eq:faber}
 \sum_{j>0\ {\rm odd}}\frac{\mathsf z_0^{-j}}{j}\bigl(W^{j/h}\bigr)_+(p)=\frac12\log\frac{p_0+p}{p_0-p},\qquad W(p_0)=\lambda_0=\mathsf z_0^h,
\end{equation}
and differentiating in $\lambda_0$ gives a Hamiltonian flow,
\begin{equation}\label{eq:dispersionless}
 \cB^{(0)}(\lambda_0)W=\{\Omega_{\lambda_0},W\},\qquad
 \Omega_{\lambda_0}(p)=\frac{2p}{(p_0^2-p^2)\,W_p(p_0)} .
\end{equation}
The coefficient of $p^{-2}$ in the variation of $W$ is proportional to $\rho$, so the insertion preserves the symmetric sector. Moreover, the factor $2p/(p_0^2-p^2)$ already shows the pairing that distinguishes D from A, since the inserted point $p_0$ and its image $-p_0$ enter on the same footing.

\subsection{Higher genus}\label{sec:taumatch}

Equation \eqref{eq:loopoperator} was designed to reproduce the classical density. Does it reproduce the reconstructed primitive at every genus? This is not automatic, because the density identity fixes the primitive only up to terms independent of $x$, and these could depend on the times in a way that spoils the comparison. The following statement settles the question on the backgrounds that matter.

\begin{proposition}[One-boundary matching]\label{prop:quantum-tau}
On a normalized ordinary-string family in the regular domain, with the generator coefficients transported along the flows and the Pfaffian times held fixed, there is a linear normalization of the tau function for which
\begin{equation}\label{eq:quantum-tau}
 \cB(\lambda)F^D_g=\Qhat_g(\lambda),\qquad g\geq1 .
\end{equation}
This fixes each $F^D_g$ up to a constant independent of the ordinary times. The equality extends to the normalized symmetric limits of proposition~\ref{thm:symmetriclimit}.
\end{proposition}

The proof, sketched in appendix~\ref{app:boundaries}, starts from the two-wave bilinear identity of the BKP hierarchy that underlies the D tau function \cite{LiuWuZhang2009,ChengHeHu2010}. Its coincident-energy limit, combined with the conserved bilinear pairing of the pencil, gives the trace primitive of section~\ref{sec:strings} including the ordering subtraction. What remains is a difference that is constant along every ordinary time, because the transported string keeps the spectral defect of the wavefunction fixed, and a linear redefinition of $\log\tau_D$ removes it without changing any field or any connected derivative of order two or higher. A useful by-product is that the ordinary flows conserve every positive-genus endpoint constant, so transporting a normalized string along the flows keeps it normalized, which is what allows the insertion to be computed on the reconstructed fields.

\subsection{Genus zero and more boundaries}\label{sec:genuszero}

At genus zero the insertion can be done by hand. Acting with \eqref{eq:dispersionless} on the disk, and choosing the disk reference so that its variation is $\partial_p\Omega_{\lambda_0}/W_p$, gives the cylinder
\begin{equation}\label{eq:cylinder}
 \omega_{0,2}(p,q)=B_D(p,q)=\left[\frac1{(p-q)^2}+\frac1{(p+q)^2}\right]\dd p\,\dd q .
\end{equation}
The first term is the familiar double pole at coincident points, and the second is its image under $q\to-q$. On a normalized string this choice of reference is not arbitrary, because the classical time dependence of the wavefunction contributes a contact term to the tau Hessian, which is the piece that completes the raw second derivative into \eqref{eq:cylinder}. Every further insertion kills the contact term, so it affects only the unstable cylinder.

One more insertion gives three boundaries, and writing $b(p,q)=(p-q)^{-2}+(p+q)^{-2}$ for the coefficient of the paired kernel, we find
\begin{equation}\label{eq:threeboundaries}
 \omega_{0,3}(p_0,p_1,p_2)=\sum_{a/\pm}\frac{W_x(a)}{W_{pp}(a)}\prod_{i=0}^{2}b(a,p_i)\,\dd p_0\,\dd p_1\,\dd p_2 ,
\end{equation}
where the sum runs over one representative of each opposite pair of ramification points. The derivation is a short residue argument, given in appendix~\ref{app:boundaries}. On an ordinary string, $\partial_py(a)=-1/W_x(a)$, so \eqref{eq:threeboundaries} is the three-boundary residue of a topological recursion \cite{EynardOrantin2007,Eynard2011} on the curve $(W,y)$ in which the Bergman kernel is replaced by the paired kernel, and we call this the \emph{mirror recursion}. Notice that the velocities $W_x(a)$ already appear here, and they will reappear at genus one.

\label{sec:moreboundaries}
Higher insertions are repeated derivatives,
\begin{equation}\label{eq:connected}
 \widetilde{\mathcal W}_k(\lambda_1,\ldots,\lambda_k)=\h^2\cB(\lambda_k)\cdots\cB(\lambda_1)\log\tau_D,
 \qquad
 \omega_{g,k}=\widetilde W_{g,k}\prod_{i=1}^kW_p(p_i)\,\dd p_i\Big|_{\lambda_i=W(p_i),\,x=\mu},
\end{equation}
with $\widetilde W_{g,k}$ the coefficient of $\h^{2g}$ in $\widetilde{\mathcal W}_k$. They are symmetric by construction because the flows commute. In practice one never needs the full quantum insertion, since the ordinary flows also commute with reconstruction, in the sense that inserting and then reconstructing gives the same answer as reconstructing and then inserting with the dispersionless flows \eqref{eq:dispersionless}. This follows from uniqueness together with the conservation of endpoint constants, and it turns each $\omega_{g,k}$ into a finite calculation with leading jets. At genus one the insertion can even be evaluated on the critical-value velocities themselves, as we will see in the next section.

Do these amplitudes satisfy a topological recursion, as the loop equations of matrix models and the W-constraints of the reduced KP hierarchies would suggest \cite{DijkgraafVerlindeVerlinde1991,Fukuma:1990jw}? On the regular normalized families three of the defining properties hold at every stable order.
\begin{proposition}[Poles and the linear loop equation]\label{prop:looplinear}
On regular normalized ordinary-string families every stable $\omega_{g,k}$, that is with $2g-2+k>0$, is a symmetric rational differential with poles only at the ramification points, regular at zero and infinity and odd under $p_i\to-p_i$. It satisfies the linear loop equation at each ramification point, which says that the sum of $\omega_{g,k}$ and its image under the exchange of the two sheets meeting there is holomorphic and vanishes at that point. Reflection parity gives the paired projection property.
\end{proposition}
Appendix~\ref{app:boundaries} sketches the proof, whose essential point is local. The two sheets that meet at a branch point have a projector sum that is holomorphic there, and this local exchange is different from the global reflection $p\to-p$. Together, the three properties of the proposition reduce the identification with the mirror recursion to a single quadratic loop equation. In local coordinates it says that a specific combination of connected amplitudes and products of lower amplitudes is holomorphic with a double zero at each ramification point. The explicit calculations agree with the mirror recursion for the one-holed torus, for the genus-one two-boundary amplitude on a regular low-rank curve and for four boundaries at genus zero, which is consistent with the quadratic equation. None of these checks proves the quadratic equation, even in these cases, and the genus-one two-boundary comparison is only numerical, so a derivation from the hierarchy is still needed.
\section{Closed surfaces}\label{sec:closed}

A closed surface has no boundary to integrate against, so its free energy has to be read from the fields, and equation \eqref{eq:taunorm} does this. The top field at genus $g$ is the second $x$-derivative of $F^D_g$, so once the reconstruction has produced $u_{n-1,g}$, the closed free energy follows up to an affine function of $x$. The interesting question is whether these free energies have a simple description in terms of the geometry of the curve. At genus one they do, and the description can be derived from the pencil.

\subsection{Genus one from the pencil}\label{sec:genusone}

A natural guess comes from semisimple Frobenius geometry, where the genus-one free energy contains one twenty-fourth of the logarithm of the product of the canonical velocities \cite{DubrovinZhang1998}. The canonical coordinates of our curve are its critical values, whose velocities are the $v_a$ of \eqref{eq:velocity}. Since opposite points have the same velocity, consider
\begin{equation}\label{eq:Fone}
 F_*=\frac1{24}\sum_{a/\pm}\log v_a ,
\end{equation}
with one representative of each of the $n$ opposite pairs. The general semisimple formula allows an additional function of the leading moduli, and there is no a priori reason for it to vanish in the moving pencil coordinates we use. Rather than assume \eqref{eq:Fone}, we therefore test it by inserting a boundary into $F_*$ and asking whether the result is the reconstructed one-boundary primitive at genus one.

The insertion of a boundary at energy $\lambda=W(p,x)$ into a critical value follows from \eqref{eq:dispersionless}, and since $W_p(a)=0$ only the $p$-derivative of $\Omega$ survives,
\begin{equation}\label{eq:insert-critical}
 \cB^{(0)}(\lambda)\,W(a,x)=v_a\,\frac{(p-a)^{-2}+(p+a)^{-2}}{W_p(p)}
 =-v_a\,\partial_\lambda\Bigl(\frac1{p-a}+\frac1{p+a}\Bigr).
\end{equation}
The velocity is the $x$-derivative of the critical value, and the insertion commutes with $\partial_x$ at fixed energy. Dividing by $v_a$ and writing $\ell_a=\dd\log v_a/\dd x$, the insertion into $\log v_a$ is $-(\ell_a+\cD)\,\partial_\lambda$ acting on the same bracket. Summing over pairs gives a compact candidate for the primitive,
\begin{equation}\label{eq:Qstar}
 24\,\cB^{(0)}(\lambda)F_*=-\partial_\lambda\Bigl(\Upsilon+\cD\,\partial_p\log T\Bigr),\qquad
 \Upsilon(p)=\sum_a\frac{\ell_a}{p-a},
\end{equation}
where both sums now run over all $2n$ roots of $T$ and $\partial_p\log T=\sum_a(p-a)^{-1}$.

Is this the reconstructed primitive? Write the genus-one problem \eqref{eq:fixedenergy} with an unknown field correction $\Phi_1$, and replace $\Upsilon$ in \eqref{eq:Qstar} by an arbitrary odd function. Then the equation $\cD\Qhat_1=\Rhat_2$ can always be solved, with
\begin{equation}\label{eq:Phi1}
 24\,\Phi_1=\partial_p\frac{W_{pp}W_{xx}-W_{px}^2}{W_p}+W_p\,\cD\Upsilon-24\,\mathfrak e,
\end{equation}
where $\mathfrak e$ collects the second $x$-derivatives of the fields together with the ordering term of \eqref{eq:rho-symbol}. This algebraic identity is proved in appendix~\ref{app:genusone}. The genus-one equation is therefore solved by a whole family of candidates, and the analytic conditions of section~\ref{sec:endpoints} decide which one is admissible, as follows.
\begin{itemize}
 \item At a ramification point $a$, the first term of \eqref{eq:Phi1} has a double pole with coefficient $-v_a'$, while $W_p\cD\Upsilon$ has a double pole with coefficient $v_a$ times the residue of $\Upsilon$. A field correction cannot have poles there, so the residue of $\Upsilon$ at $a$ must be $v_a'/v_a=\ell_a$, and this is where the logarithm comes from.
 \item At the origin, both the first term and $\mathfrak e$ have fourth-order poles, which cancel because of the term $-(\rho')^2$ in the ordering correction \eqref{eq:rho-symbol}, so the genus-one free energy probes how the nonlocal term is ordered.
 \item At infinity, the growth of $\Phi_1$ is that of a field correction, with a top coefficient equal to $h/24$ times $\sum_a\ell_a'$.
\end{itemize}
Uniqueness now identifies the candidate with the reconstruction, and \eqref{eq:taunorm} turns its top coefficient into the free energy. Comparing first derivatives through proposition~\ref{prop:quantum-tau}, whose proof we sketch, removes the remaining ambiguity along the ordinary times.

\begin{proposition}[Genus one on a moving background]\label{prop:moving-genus-one}
On every moving D curve satisfying \eqref{eq:regular}, the genus-one reconstruction obeys
\begin{equation}\label{eq:genusone}
 u_{n-1,1}=2h\,\partial_x^2F_*,\qquad \Qhat_1=\cB^{(0)}F_* .
\end{equation}
On a normalized ordinary string, $F_*$ equals $F^D_1$ up to a constant independent of all ordinary times.
\end{proposition}

The same three steps work for the scalar operator of Part I, where the polynomial $T$ is replaced by $W_p$, there is no origin, and the ordering term is that of a differential operator. The result is the A-series analogue
\begin{equation}\label{eq:FoneA}
 F^A_1=\frac1{24}\sum_{a=1}^{h-1}\log v_a ,
\end{equation}
summed over the $h-1$ ramification points of a scalar curve with $h$ sheets. Appendix~\ref{app:genusone} derives the top-field identity in full, but the first derivatives of $F^A_1$ along the ordinary times use the scalar counterpart of proposition~\ref{prop:quantum-tau}, which we do not write out. Both series therefore obey the same genus-one law and differ only in how many canonical directions they count.

\label{sec:fixedpoint}
What happens to \eqref{eq:Fone} when the Pfaffian field is turned off? Near the limit the two origin directions come from the roots $t=\pm\rho_0/\sqrt{A'(0)}+O(\rho_0^2)$ of $t^2A'(t)=\rho_0^2$. Their velocities are
\begin{equation}\label{eq:originvelocities}
 v_\pm=\partial_xA(0)\pm2\sqrt{A'(0)}\,\partial_x\rho_0+O(\rho_0).
\end{equation}
Along a normalized string the origin equation \eqref{eq:origin} gives $\rho_0=\gamma_0/\pi_p(0)$ at leading order, so $\partial_x\rho_0$ vanishes with $\gamma_0$. Both velocities therefore tend to the velocity of the fixed critical point $p=0$,
\begin{equation}\label{eq:vfix}
 v_{\rm fix}=W_x(0,x).
\end{equation}
The other $n-2$ pairs are ordinary critical points of the symmetric curve, and comparing with the A-series sum \eqref{eq:FoneA} for the same even energy, in which each nonzero pair is counted twice and the origin once, we find
\begin{equation}\label{eq:F1fixed}
 F^D_1=\frac12F^A_1+\frac1{16}\log v_{\rm fix}
\end{equation}
on ordinary deformations of the symmetric curve. In words, the D free energy at genus one is half the A free energy plus a term that lives entirely at the origin.

On a homogeneous background every critical value scales with the same power of $x$, and so does every velocity. Each logarithm then contributes the same coefficient of $\log x$, so the ratio of the two genus-one free energies is a ratio of counts,
\begin{equation}\label{eq:R1}
 R_1=\frac{F^D_1}{F^A_1}=\frac{n}{h-1},
\end{equation}
which depends neither on the partner degree nor on the details of the background.

\subsection{Higher genus}\label{sec:highergenus}

At genus two and beyond we do not have a derivation, but the reconstruction produces exact data on which to test formulas. We use homogeneous backgrounds, on which all scales are powers of $x$. With $H=h+s$ and $\mathfrak d=(h+s-1)/2$, the leading energy takes the form $W=x^{h/(2\mathfrak d)}w(z)$ with $z=p\,x^{-1/(2\mathfrak d)}$, and the vector free energy $F_g^{\vecc}=2F^D_g$ is a single number times $x^{-(g-1)H/\mathfrak d}$. Within this class the reconstructed top field fixes that number uniquely.

At genus two the critical points acquire weights. For a nonzero critical value we define the log-Hessian $\vartheta_a=w''(a)/w(a)$, and the exact data are then reproduced by the following formula.
\begin{conjecture}[Genus two]\label{conj:g2}
On regular homogeneous backgrounds,
\begin{equation}\label{eq:g2law}
 F_2^{\vecc}=x^{-H/\mathfrak d}\,\widetilde\Phi_2(h,s)\sum_a\vartheta_a,\qquad
 \widetilde\Phi_2=-\frac{(2s-h)(2h+s)(h+2s)}{5760\,h\,(h+s-1)^2},
\end{equation}
\end{conjecture}
where the sum runs over all $2n$ momentum critical points. The coefficient has a local origin, since it is the Witten--Kontsevich genus-two free energy evaluated on a power-law solution, dressed by a Jacobian that makes the local kinetic term constant \cite{Witten1991,Kontsevich1992,ItzyksonZuber1992}. In the symmetric limit the origin again needs care, because each of the four points that approach $p=0$ carries four times the polynomial log-Hessian of the origin, by \eqref{eq:cluster}. At a Chebyshev endpoint this gives a closed ratio of the D and A genus-two free energies, which we record in section~\ref{sec:potts}.

At genus three, however, the dependence on the background shows that a sum over local terms is no longer enough. One also needs to know how the Hessian at one critical point changes when another critical value is moved. Such terms are familiar from semisimple higher-genus reconstruction \cite{Givental2001,Teleman2012,DuninBarkowskiEtAl2014}. Appendix~\ref{app:homogeneous} gives the resulting formula, which involves two further coefficient functions besides the local Witten--Kontsevich term and agrees with every reconstructed field we have computed. We state the genus-two and genus-three formulas as conjectures, and it would be interesting to derive them in the way section~\ref{sec:genusone} derives the genus-one law.
\section{Potts and tetracritical Ising}\label{sec:potts}

We now put the recipe to work on the example that motivated it. The tetracritical Ising model and the three-state Potts model are the $(A_4,A_5)$ and $(A_4,D_4)$ minimal models, which both have central charge $c=4/5$ and are both labelled by $(p,p')=(6,5)$. They differ in their modular invariant \cite{CappelliItzyksonZuber1987}, which for Potts keeps only part of the diagonal spectrum and adds a non-diagonal sector. In the minimal string this sector supplies an extra tachyon, which we will see as the Pfaffian coupling. In the language of the previous sections we have $h=6$, $n=4$ and partner degree $s=5$.

\label{sec:pottsfamily}
Both theories can be placed on the same homogeneous moving family,
\begin{equation}\label{eq:pottsfamily}
 W(p,x)=p^6-6x^{1/5}p^4+9x^{2/5}p^2-2x^{3/5},\qquad
 \pi(p,x)=p^5-5x^{1/5}p^3+5x^{2/5}p,
\end{equation}
with endpoint $\mu=1$. In the scaling variable $z=p\,x^{-1/10}$ the energy is $W=x^{3/5}e_6(z)$, where $e_k$ denotes the Chebyshev polynomial with $e_k(2\cos\theta)=2\cos k\theta$, so at the endpoint we sit at the conformal point $W=e_6$, $\pi=e_5$. A direct computation gives $\{\pi,W\}=-6$, so $\kappa=-6$ and the two theories share the disk $y=\pi/6$.

The quantum operators, however, are different. For tetracritical Ising we use the scalar operator of Part I, while for Potts we use the pencil,
\begin{align}\label{eq:pottsops}
 Q_A&=D^6+\frac12\{u^A_4,D^4\}+\frac12\{u^A_2,D^2\}+u^A_0, & Q_A\psi&=\lambda\psi,\\
 L_D&=D^7+\frac12\{u_3,D^5\}+\frac12\{u_2,D^3\}+\frac12\{u_1,D\}, & L_D\psi&=\lambda D\psi .\nonumber
\end{align}
The Potts solution is the invariant one, with $\rho=0$ at every genus. The two sets of leading fields agree, but the operators order them differently, and so their quantum corrections already differ at genus one.

The ramification points are the zeros of $e_6'(z)=6z(z^2-1)(z^2-3)$. For tetracritical Ising there are five of them, $z=0,\pm1,\pm\sqrt3$, each a canonical direction. For Potts the nonzero points form two opposite pairs, and the origin, reached from $\rho\neq0$, carries two canonical directions, so Potts has four canonical directions against five. The two-variable potential \eqref{eq:LG} makes the count visible,
\begin{equation}\label{eq:pottsLG}
 \cF(t,\mathsf y)=t^3-6t^2+9t-2-t\mathsf y^2+2\rho\mathsf y .
\end{equation}
At $\rho=0$ its critical points are $(t,\mathsf y)=(0,\pm3)$, $(1,0)$ and $(3,0)$, all of them Morse, and the first two share the critical value $-2$. Switching on $\rho$ splits that value into $-2\pm6\rho$.

\subsection{Genus one}\label{sec:pottsg1}

At genus one the reconstruction reduces to linear algebra, because by homogeneity the field correction and the primitive take the forms
\begin{equation}\label{eq:pottsg1ansatz}
 \Phi_1=\phi_{1,0}\,x^{-8/5}+\phi_{1,2}\,x^{-9/5}p^2+\phi_{1,4}\,x^{-2}p^4,\qquad
 \Qhat_1=x^{-17/10}\,\frac{N(z^2)}{720\,z^5(z^2-1)^5(z^2-3)^5},
\end{equation}
with a numerator $N$ of degree nine in $z^2$. The denominator places poles of order five at the five ramification points, and the numerator degree enforces decay at infinity. Clearing the denominator in \eqref{eq:fixedenergy} gives fifteen linear equations for these thirteen unknowns. In both theories the system has rank thirteen and the two extra equations are satisfied, which is the existence statement of section~\ref{sec:strings} at work. The recovered fields are listed in table~\ref{tab:pottsg1}, and the resulting one-holed tori are
\begin{equation}\label{eq:pottsW11}
 W_{1,1}^{A}(z)=-\frac{\mathcal N_A(z)}{120\,z^4(z^2-1)^4(z^2-3)^4},\qquad
 W_{1,1}^{\rm Potts}(z)=-\frac{\mathcal N_D(z)}{30\,z^4(z^2-1)^4(z^2-3)^4},
\end{equation}
with the two numerators recorded in appendix~\ref{app:potts}. Both primitives decay as $z^{-7}$ at infinity, as \eqref{eq:endpoints} requires, and both one-holed tori have a pole at the origin, which is a ramification point of the symmetric curve. Their difference is short,
\begin{equation}\label{eq:pottsdiff}
 W_{1,1}^{\rm Potts}-W_{1,1}^{A}=-\frac{2z^{10}-13z^8+25z^6-8z^4-11z^2+9}{40\,z^4(z^2-1)^2(z^2-3)^2}.
\end{equation}
It has poles at the nonzero critical points as well as at the origin, so the origin does not account for the whole difference.

\begin{table}[t]
\centering
\begin{tabular}{@{}lccccc@{}}
\toprule
& $\phi_{1,0}$ & $\phi_{1,2}$ & $\phi_{1,4}$ & $F_1$ & directions\\
\midrule
Tetracritical Ising & $57/100$ & $-21/10$ & $1/2$ & $-\tfrac1{12}\log x$ & $5$\\
Potts & $66/25$ & $-108/25$ & $4/5$ & $-\tfrac1{15}\log x$ & $4$\\
\bottomrule
\end{tabular}
\caption{Genus-one field corrections \eqref{eq:pottsg1ansatz} on the common family \eqref{eq:pottsfamily}, and the free energies obtained from the top coefficient $\phi_{1,4}$ through $u^A_{4,1}=6\partial_x^2F_1^A$ and \eqref{eq:taunorm}. The last column counts canonical directions. Both free energies agree with the velocity laws \eqref{eq:FoneA} and \eqref{eq:Fone}.}\label{tab:pottsg1}
\end{table}

The free energies in table~\ref{tab:pottsg1} illustrate section~\ref{sec:closed}. All critical values of \eqref{eq:pottsfamily} scale as $x^{3/5}$, so every velocity scales as $x^{-2/5}$ and every logarithm in \eqref{eq:Fone} or \eqref{eq:FoneA} contributes the same multiple of $\log x$. The ratio of the two free energies is therefore the ratio of the numbers of canonical directions, in agreement with \eqref{eq:R1}. In the A sum each of the two nonzero pairs is counted twice and the origin once, while in the D sum each pair is counted once and the origin twice. Put differently, the Potts free energy is half the tetracritical free energy plus the origin term of \eqref{eq:F1fixed}.

Where does the origin enter $W_{1,1}$ itself? In a recursion on the curve each canonical direction contributes a residue, and the nonzero ramification points of Potts contribute like any other branch point. The resolved origin is different, because four momentum points collide there. Evaluating the local residue of the mirror recursion on the collision and keeping the first subleading coefficients of the disk and of the energy, we find that the origin contributes
\begin{equation}\label{eq:pottsdoublet}
 -\left(\frac1{30\,p^4}+\frac1{18\,p^2}\right)\dd p
\end{equation}
to the Potts one-holed torus, and adding the two nonzero representatives reproduces \eqref{eq:pottsW11} exactly. The inverse-square term in \eqref{eq:pottsdoublet} is not visible in the naive collision model. It comes from the subleading local data, which have to be kept alongside the leading monomials, as the calculation in appendix~\ref{app:potts} shows.

\subsection{Higher genus}\label{sec:pottshigher}

Beyond genus one the homogeneous reconstruction becomes a three-by-three linear problem at each genus. The new fields scale as $x^{(3-11g)/5}$, $x^{(2-11g)/5}$ and $x^{(1-11g)/5}$, and the same matrix acts in both theories, with determinant $\mathsf m(\mathsf m-1)(\mathsf m-2)$ for $\mathsf m=11g-3$, so only the sources differ. Appendix~\ref{app:potts} lists them together with the fields through genus three, and the free energies follow from the top fields, with their ratios collected in table~\ref{tab:pottsratios}.

\begin{table}[t]
\centering
\begin{tabular}{@{}cccc@{}}
\toprule
$g$ & $F_g^A$ & $F_g^D$ & $R_g=F_g^D/F_g^A$\\
\midrule
$1$ & $-\tfrac1{12}\log x$ & $-\tfrac1{15}\log x$ & $4/5$\\[2pt]
$2$ & $\tfrac{119}{3600}\,x^{-11/5}$ & $\tfrac{17}{450}\,x^{-11/5}$ & $8/7$\\[2pt]
$3$ & $\tfrac{1180361}{10800000}\,x^{-22/5}$ & $\tfrac{13124}{84375}\,x^{-22/5}$ & $98816/69433$\\
\bottomrule
\end{tabular}
\caption{Closed free energies of tetracritical Ising and Potts on the family \eqref{eq:pottsfamily}, from exact reconstruction through genus three. The genus-one and genus-two ratios agree with \eqref{eq:R1} and \eqref{eq:R2}. Appendix~\ref{app:homogeneous} gives the same ratios for further D-series backgrounds.}\label{tab:pottsratios}
\end{table}

The log-Hessian law \eqref{eq:g2law} explains the genus-two ratio. At a Chebyshev endpoint the nonzero critical points have $\vartheta_j=-(h^2/4)\csc^2(\pi j/h)$. The A sum contains the origin once with $\vartheta_{\rm fix}=-h^2/4$, while the D sum contains four points near the origin, each with four times that value. Replacing $\vartheta_{\rm fix}$ by $16\vartheta_{\rm fix}$ and halving the vector sum gives
\begin{equation}\label{eq:R2}
 R_2=\frac12\left[1+\frac{15}{\sum_{j=1}^{h-1}\csc^2(\pi j/h)}\right]=\frac{h^2+44}{2(h^2-1)},
\end{equation}
which at $h=6$ is the value in table~\ref{tab:pottsratios}. The genus-three ratio has no such closed form, because the genus-three formula also involves the motion of one Hessian under a change of another critical value. It nevertheless passes an independent test. The height-model construction of Kostov and of Higuchi and Kostov gives the unitary vacuum amplitudes of both theories as sums over graphs \cite{Kostov1992,HiguchiKostov1995}. The combination $R_3R_0^2$, which is independent of how genus is counted ($R_0=F_0^D/F_0^A$ is the genus-zero ratio), agrees with the graph sum, as the numbers in appendix~\ref{app:homogeneous} show.

\subsection{Boundaries and the worldsheet}\label{sec:pottsboundaries}

The cylinder is where the pairing of the D series is most visible. For tetracritical Ising the cylinder on the curve \eqref{eq:pottsfamily} is the Bergman kernel of the $z$-plane, $\dd z_1\dd z_2/(z_1-z_2)^2$, whereas for Potts it is the paired kernel \eqref{eq:cylinder}. Its second term is the image contribution, which comes from the factor $2p/(p_0^2-p^2)$ in the insertion \eqref{eq:dispersionless}. The three-boundary amplitude \eqref{eq:threeboundaries} then pairs every external point with its image at every ramification point.

At genus one, two ordinary insertions of the free energy give the two-boundary amplitude. On the Potts family it takes the form
\begin{equation}\label{eq:pottsW12}
 \omega_{1,2}^{\rm Potts}=\frac{\mathcal N_{12}(z_1^2,z_2^2)\,\dd z_1\,\dd z_2}{75\,z_1^6z_2^6\,\mathcal C(z_1^2)^6\,\mathcal C(z_2^2)^6},\qquad \mathcal C(Z)=(Z-1)(Z-3),
\end{equation}
with a symmetric polynomial numerator of degree fourteen in each variable, whose coefficients are listed in appendix~\ref{app:potts}. Its sixth-order poles are the ones allowed for a genus-one amplitude with two boundaries, and they sit only at the ramification points.

\label{sec:pottsbulk}
Finally we turn to the worldsheet. Rodriguez, Usatyuk and Wang computed torus one-point amplitudes of diagonal tachyons in both theories by direct integration over the moduli space of the torus \cite{RodriguezUsatyukWang2025}, and to compare with them we need a coupling frame. A diagonal tachyon with odd labels $(r,s')$ has index $m=|sr-hs'|$. When $0<m<h$ and $s$ does not divide $m$ we call it primary, and we identify its coupling with the ordinary time of index $m$, up to the normalization fixed by the sphere three-point function. The identity $(1,1)$ fixes one overall constant per model. The torus one-point amplitude is then the derivative of the genus-one free energy along the corresponding flow, evaluated at the conformal point.

For $(h,s)=(6,5)$ two tachyons are primary, and table~\ref{tab:pottsT1} shows that our values agree exactly with the worldsheet numbers in both theories. The table also shows how the origin enters. By \eqref{eq:F1fixed}, the Potts amplitude is half the tetracritical one plus the derivative of $\log v_{\rm fix}/16$, which is the last column of the table.

\begin{table}[htb]
\centering
\begin{tabular}{@{}cccccc@{}}
\toprule
$(r,s')$ & $m$ & $T^A$ & $T^{D}$ & $T^D-\tfrac12T^A$ & worldsheet \cite{RodriguezUsatyukWang2025}\\
\midrule
$(1,1)$ & $1$ & $20$ & $16$ & $6$ & $20$, $16$\\
$(3,3)$ & $3$ & $6$ & $0$ & $-3$ & $6$, $0$\\
\bottomrule
\end{tabular}
\caption{Torus one-point amplitudes of the primary diagonal tachyons at $(h,s)=(6,5)$, for tetracritical Ising ($T^A$) and Potts ($T^D$). The column $T^D-\tfrac12T^A$ is the origin contribution of \eqref{eq:F1fixed}. The identity row fixes the normalization of each model and is not a test. The count \eqref{eq:R1} is tested instead by the vacuum free energies of table~\ref{tab:pottsratios}. The worldsheet column lists the A and D values. The general formula and all other models are in appendix~\ref{app:worldsheet}.}\label{tab:pottsT1}
\end{table}

The other labels of the worldsheet tables have $m>h$ and couple to gravitational descendants in the hierarchy frame, and for $m>h+s$ the worldsheet integrals also pick up residues from poles that cross the Liouville contour \cite{RodriguezUsatyukWang2025}. We do not attempt these labels here. Within the primary range, however, the comparison goes beyond individual entries. Appendix~\ref{app:worldsheet} sketches why, throughout that range, the derivatives of \eqref{eq:Fone} and \eqref{eq:FoneA} reproduce the closed formulas that ref.~\cite{RodriguezUsatyukWang2025} conjectured from their numerics, for both series and for every coprime pair with $s<2h$, with $h$ even in the D case.

The Pfaffian tachyon gives one more observable. Its coupling is the origin constant $\gamma_0$, which the first Pfaffian flow moves at the rate \eqref{eq:origintangent} at the symmetric point. The ordinary insertions leave the dependence of the free energy on $\gamma_0$ undetermined up to an additive function, so the second derivative in $\gamma_0$ needs one more input, an identity for the Pfaffian Hessian that we state as a conjecture in appendix~\ref{app:worldsheet}. Granting it, the second derivative is a pure Pfaffian torus two-point amplitude. For Potts it is then a prediction, recorded with its normalization in appendix~\ref{app:worldsheet} together with closed formulas for partner degrees three, five and seven. At genus zero, and for labels in the primary range, the same coupling reproduces the mixed sphere four-point amplitudes of the D series computed in ref.~\cite{RodriguezUsatyukWang2025}, once the operator normalizations have been fixed by three-point functions.

Let us return to the question we started with. Potts and tetracritical Ising share their central charge and their conformal curve, and each difference we have found between their amplitudes traces back to how the D theory treats the origin and the pairing $p\to-p$. This is most explicit in the genus-one identity \eqref{eq:F1fixed}, in which the origin term $\log v_{\rm fix}/16$ is all that separates the Potts free energy from half the tetracritical one. In the torus one-point amplitudes the same term gives the last column of table~\ref{tab:pottsT1}. In the one-holed torus the origin contributes the residue \eqref{eq:pottsdoublet}, but the difference \eqref{eq:pottsdiff} also has double poles at the nonzero critical points, which no residue at the origin can produce. In the cylinder, finally, the difference is the image term of the paired kernel.

\section{Discussion}\label{sec:discussion}

We set out to see whether the differential-equation approach to boundary amplitudes, which Johnson developed for the KdV hierarchies and which we extended to the reduced KP hierarchies in Part I, survives in the D series. It does, and what makes it work is the second pole of the classical energy, which the nonlocal Pfaffian term of the pencil places at zero momentum. Asking the integrated resolvent to be regular at both poles fixes every quantum correction, and the condition at the origin removes the one ambiguity that the condition at infinity leaves behind, namely the first Pfaffian flow. Ordinary string equations supply a solution, as we have sketched, and the solutions continue analytically to the symmetric sector where the Pfaffian field vanishes.

Boundaries are added in the same spirit as in Johnson's work, by combining the ordinary odd flows of the D hierarchy into a loop operator, which reproduces the reconstructed primitive at every genus. At genus zero the loop operator gives a cylinder that pairs every point with its image under $p\to-p$, together with the three-boundary residue of a mirror recursion. At higher order the amplitudes have the poles and the linear loop equations of that recursion, while the quadratic loop equation remains open. At genus one the closed free energy follows from the pencil itself as one twenty-fourth of the logarithm of the product of critical-value velocities. The same argument gives the A-series law, up to the scalar tau matching, so the two series differ at genus one only in how many canonical directions they count.

This counting is what separates Potts from tetracritical Ising. The two theories share their central charge and their conformal spectral curve, and at the conformal point the D curve has no Pfaffian field at all, so all that survives of it is the resolved origin. Since the D reconstruction reaches the symmetric sector from the regular one, it remembers that the critical point at zero momentum carries two canonical directions. We have followed this memory through the free energies to genus three, the one-holed torus and the torus one-point amplitudes of the worldsheet, and the pairing $p\to-p$ through the cylinder. In the primary range, the genus-one law reproduces the torus one-point formulas conjectured from worldsheet numerics, for both series and for every coprime pair with $s<2h$, and the Pfaffian coupling reproduces the mixed sphere four-point amplitudes. Conditional on an identity for the Pfaffian Hessian, we also predict pure Pfaffian torus two-point amplitudes.

Several questions now have a concrete form.
\begin{itemize}
 \item \emph{The quadratic loop equation.} We have sketched why the connected amplitudes generated by the loop operator have the pole structure, the linear loop equation and the projection property of a topological recursion with the paired kernel. The remaining quadratic loop equation would complete the identification with the mirror recursion at every stable order. The abstract loop-equation framework \cite{BorotEynardOrantin2015} and the odd recursion of Alexandrov and Shadrin \cite{AlexandrovShadrin2023} provide useful comparisons.
 \item \emph{Higher-genus laws.} The log-Hessian formula at genus two and its refinement at genus three reproduce every reconstructed field we have computed, but they remain conjectures. Can the genus-one derivation of section~\ref{sec:genusone} be extended? The relation between the D hierarchy and its Dubrovin--Zhang counterpart \cite{Wu2013}, the semisimple genus-two formulas \cite{DubrovinLiuZhang2012} and the local structure of higher-genus free energies \cite{ShadrinWang2026} are natural tools, as are the W-constraints and the recursion for simple singularities \cite{BakalovMilanov2013,Milanov2015,VanDeLeur2014}.
 \item \emph{The second family of flows.} Our boundaries use only the ordinary flows, but the Pfaffian flows generate other insertions. Matching the expansion at the second energy pole to the second root of the Lax operator would give a new kind of boundary and mixed ordinary and Pfaffian amplitudes. The first step is the Pfaffian Hessian of conjecture~\ref{conj:pfaffian}, which would fix the dependence of the free energy on the origin constant. The matrix-resolvent formulas for Drinfeld--Sokolov tau functions offer an independent framework for comparison \cite{BertolaDubrovinYang2021}.
 \item \emph{Beyond the primary range.} The worldsheet tables contain many tachyons with index $m>h$, which couple to gravitational descendants in the hierarchy frame, and those with $m>h+s$ also pick up pole-crossing residues on the worldsheet. A complete coupling dictionary, presumably organized by the resonance transformations of minimal gravity \cite{BelavinZamolodchikov2009,BelavinDubrovinMukhametzhanov2014}, would allow us to reach them, to test our conditional Pfaffian torus two-point predictions, and to match the loop insertion to definite boundary conditions of the matter theory \cite{Cardy1989,FateevZamolodchikovZamolodchikov2000,Teschner2000,SeibergShih2004,AleshkinBelavin2019,RodriguezUsatyukWang2025}.
 \item \emph{The E series and matrix models.} The E-series minimal strings are the remaining part of the ADE family, and the worldsheet data of ref.~\cite{RodriguezUsatyukWang2025} include them. More broadly, the D and E minimal strings are expected to be dual to multi-matrix integrals that have not been solved. The reconstruction gives their perturbative observables without the matrices, and it would be interesting to know whether it can also guide the search for the matrix models.
 \item \emph{Nonperturbative physics and singular limits.} The string equation and the resolvent equation both contain information beyond perturbation theory, as emphasized in ref.~\cite{AhmedJohnsonSaraswat2025}, and in the D series this information includes the Pfaffian sector. Instanton calculations in matrix models show what kind of information to expect \cite{MarinoSchiappaWeiss2008}. In a different direction, more singular limits than the one studied here, where the hodograph Jacobian degenerates or more critical points collide, would require new local scalings. The general theory of limits of topological recursion offers criteria to test them against \cite{BouchardEynard2013,BorotEtAl2025Limits}.
\end{itemize}

\acknowledgments
J.N. thank the organizers of ``Holography and Observables in Quantum Gravity'' at RIKEN, Japan, where part of this work was done. We are grateful to Mykhaylo Usatyuk and  Kazumi Okuyama for useful discussions. Claude and GPT Astra were used in developing symbolic calculations, consistency checks and manuscript preparation. The scope of the computational checks is stated with the results. W.A. is supported by the U.S. Department of Energy under grant DE-SC 0011702. J.N. is supported by the Beijing Institute of Mathematical Sciences and Applications.

\appendix
\section{Symbols, ordering and residues}\label{app:conventions}

This appendix collects the conventions behind section~\ref{sec:density}. Symmetrically ordered coefficients have Weyl symbols
\begin{equation}\label{eq:weylorder}
 \operatorname{symb}\Bigl(\tfrac12\{f,D^j\}\Bigr)=\sum_{\ell=0}^{\lfloor j/2\rfloor}\frac{\binom j{2\ell}}{2^{2\ell}}\,\h^{2\ell}f^{(2\ell)}p^{j-2\ell},
\end{equation}
so the leading symbol replaces $D$ by $p$ and every correction carries an even number of $x$-derivatives. The nonlocal term keeps its order, which gives \eqref{eq:rho-symbol}. The formal inverse $D^{-1}$ obeys $D^{-1}f=\sum_{k\geq0}(-\h)^kf^{(k)}D^{-1-k}$ in the convention $D=\h\partial_x$.

Write $\mathsf A_{\rm W}=\sum_k\h^{2k}\mathsf A_{2k}$ and $G=\sum_l\h^{2l}G_{2l}$. Averaging the left and right inverse equations removes the odd orders of the star product and gives the recurrence
\begin{equation}\label{eq:inverserecurrence}
 G_0=\frac1{\mathsf A_0},\qquad
 G_{2g}=-\frac1{\mathsf A_0}\sum_{\substack{k+l+\ell=g\\(k,\ell)\neq(0,0)}}\mathcal M_{2\ell}(\mathsf A_{2k},G_{2l}),
\end{equation}
with the bidifferential operators
\begin{equation}\label{eq:bidiff}
 \mathcal M_{2\ell}(f_1,f_2)=\frac1{2^{2\ell}(2\ell)!}\sum_{j=0}^{2\ell}(-1)^j\binom{2\ell}{j}\bigl(\partial_p^{2\ell-j}\partial_x^jf_1\bigr)\bigl(\partial_p^j\partial_x^{2\ell-j}f_2\bigr).
\end{equation}
The first instance is \eqref{eq:G2}. All derivatives are taken at fixed $\lambda$, and the sheet relation is imposed only afterwards.

At a sheet root $W(p_\lambda,x)=\lambda$, a Taylor expansion in $W-\lambda$ gives the residue rule
\begin{equation}\label{eq:residuerule}
 \Res_{p=p_\lambda}\frac{f(p,x)\,\dd p}{(W-\lambda)^j}=\frac1{(j-1)!}\,\partial_\lambda^{\,j-1}\Bigl(\frac f{W_p}\Bigr)\Big|_{p=p_\lambda},\qquad \partial_\lambda=\frac1{W_p}\partial_p .
\end{equation}
The derivatives hold $x$ and $\lambda$ fixed. Applied to \eqref{eq:G2} with $\mathsf A_0=p(W-\lambda)$ and $\mathsf A_2=p\Phi_1+O_D$, where
\begin{equation}\label{eq:aord}
 O_D=\sum_{i=1}^{n-1}\frac{(2i-1)(2i-2)}8\,u_{i,0}''\,p^{2i-3}+\frac{\rho_0\rho_0''-(\rho_0')^2}{2p^3}
\end{equation}
collects the ordering terms, this produces the full genus-one density in a form that can be evaluated directly on any moving curve. Appendix~\ref{app:genusone} uses it in exactly this way.

Two parities are used throughout. Reversing $\h$ reverses the star product, and uniqueness of the two-sided inverse makes $G$ even in $\h$. Reversing $p$ reverses the symbol of the pencil, so the densities on opposite sheets have opposite signs and every $\Rhat_{2g}$ is odd in $p$. The field variation $\Phi_g$ is even, which is used repeatedly in uniqueness arguments.
\section{Existence, normalization and the symmetric limit}\label{app:reconstruction}

Section~\ref{sec:strings} claimed that a normalized ordinary string solves the reconstruction problem. Here we explain the three ingredients behind this claim, a primitive supplied by the string, the growth of that primitive at the two poles, and the integration constants that remove what is not allowed.

\subsection{A primitive from the string and its growth}

Introduce one auxiliary component, $\chi=D^{-1}(\rho\psi)$, so that $D\chi=\rho\psi$. The pencil then becomes a first-order system
\begin{equation}\label{eq:localY}
 Y=(\psi,D\psi,\ldots,D^h\psi,\chi)^t,\qquad \h\partial_xY=\mathsf LY,
\end{equation}
of dimension $h+2=2n$. The only entry of $\mathsf L$ that depends on the energy is $\lambda$ in row $h+1$ and column $2$. Let $M$ be the normalized rank-one projector onto one sheet, written $M(p)$ when it is the sheet through the point $p$. Eliminating the auxiliary rows of the matrix inverse gives
\begin{equation}\label{eq:projectordensity}
 \Rhat=M_{2,h+1}-\frac\h2\,\partial_xM_{1,h+1},
\end{equation}
because the $(1,h+1)$ entry of the inverse is the scalar inverse $G$ and the $(2,h+1)$ entry is $p\star G$.

The string equation provides a connection in $\lambda$. Differentiating $\cL\psi=\lambda\psi$ and using $[\cL,S]=-\kappa\h$ gives $\kappa\h\,\partial_\lambda\psi=-S\psi$. In the basis \eqref{eq:localY} this becomes a polynomial matrix $\mathsf B(\lambda)$ obeying the compatibility condition
\begin{equation}\label{eq:spectralconnection}
 \partial_x\mathsf B+\frac{[\mathsf B,\mathsf L]}\h=\partial_\lambda\mathsf L .
\end{equation}
The origin equation \eqref{eq:origin} is what makes $\mathsf B$ local. Cyclicity of the trace and $\h\partial_xM=[\mathsf L,M]$ then give $\partial_x\tr(\mathsf BM)=M_{2,h+1}$, and comparing with \eqref{eq:projectordensity} yields the corrected trace primitive, whose second term is the ordering subtraction of section~\ref{sec:strings},
\begin{equation}\label{eq:correctedtrace}
 \Qhat=\tr(\mathsf BM)-\frac\h2M_{1,h+1},\qquad \partial_x\Qhat=\Rhat .
\end{equation}
Its even part supplies the genus expansion.

Combining the scalar expression for the string primitive with the localized WKB recursion and completeness of the projectors, one finds for every $g\geq1$
\begin{equation}\label{eq:densitydecay}
 \Rhat_{2g}=\frac{u_{n-1,g}}{h^2}\,p^{-h-1}+O\bigl(p^{-h-3}\bigr)\ \ (p\to\infty),\qquad
 \Rhat_{2g}=-\frac{\rho_g}{2\rho_0^3}\,p^3+O\bigl(p^5\bigr)\ \ (p\to0).
\end{equation}
The estimates are proved by induction, first at infinity, then for the auxiliary projector entries at the origin, and finally for the full density at the origin. The step at infinity is a degree count. The origin needs more care, and we sketch it here.

Divide the pencil equation and $D\chi=\rho\psi$ by $\psi$. With $\mathsf w=D\psi/\psi$ and $\mathsf f=\chi/\psi$ they become
\begin{equation}\label{eq:localized}
 \frac{L_D^{\rm diff}\psi}{\psi}-\lambda\mathsf w+\rho\,\mathsf f=0,\qquad \mathsf w\,\mathsf f+\h\cD\mathsf f=\rho,
\end{equation}
where $L_D^{\rm diff}$ is the differential part of $L_D$. At each positive order in $\h$ the new coefficients $(\mathsf w_m,\mathsf f_m)$ solve a linear system with matrix and inverse
\begin{equation}\label{eq:originmatrix}
 \mathsf N=\begin{pmatrix}\mathsf k&\rho_0\\ \rho_0/p&p\end{pmatrix},\qquad
 \mathsf N^{-1}=\frac pT\begin{pmatrix}p&-\rho_0\\ -\rho_0/p&\mathsf k\end{pmatrix},\qquad \mathsf k=pW_p+\frac{\rho_0^2}{p^2},
\end{equation}
so the only momentum denominators are $p$ and $T$. Since $W_x/W_p=O(p)$ at zero, $\cD$ does not lower the order there.

The induction starts from the auxiliary diagonal entry $M_{\chi\chi}$ of the sheet projector, whose leading value $-\rho_0^2/T$ tends to $1/2$ at the origin. On each of the $h$ large sheets this entry is $O(p^{-h-2})$, and their sum has only integer powers of $\lambda^{-1}$, so it is $O(\lambda^{-2})$. Completeness of the projectors and the equality of the entries on the two small sheets then give $M_{\chi\chi}=\tfrac12+O(p^4)$ on the small sheets, with the constant entirely at genus zero. Two exact identities control the odd orders,
\begin{equation}\label{eq:oddidentities}
 \mathsf w+\widetilde{\mathsf w}=-\h\cD\log\Sigma,\qquad \rho\,(\mathsf f+\widetilde{\mathsf f})=\h\,\Sigma\,\cD M_{\chi\chi},
\end{equation}
where tildes denote the adjoint quantities and $\Sigma$ is the bilinear concomitant, with $\Sigma_0=T/p^2$. Together with $\cD M_{\chi\chi}=O(p^4)$ they give $\mathsf w_m,\mathsf f_m=O(p^2)$ at odd orders. At an even order $2g$ the right side of the linear system is $\rho_g(-\rho_0/p,1)^t$ plus terms of orders $(p,p^2)$, which follow from the previous bounds. Applying $\mathsf N^{-1}$ gives
\begin{equation}\label{eq:originWKB}
 \mathsf w_{2g}=\frac{\rho_g}{\rho_0}\,p+O(p^3),\qquad \mathsf f_{2g}=O(p).
\end{equation}
The positive-order coefficients of the concomitant are then bounded, so $(\Sigma^{-1})_m=O(p^4)$, and the projector density \eqref{eq:projectordensity} gives the estimate at the origin in \eqref{eq:densitydecay}. The dependency is triangular. An odd order uses known concomitant coefficients, and an even order uses the preceding coefficients and then fixes its own concomitant coefficient. The local energy coordinates $\mathsf z=\lambda^{1/h}$ at infinity and $\zeta=\lambda^{-1/2}$ at zero are invariant under $\cD$. In these coordinates the growth bound on the primitive leaves only the modes
\begin{equation}\label{eq:forbidden}
 \mathsf z^{-1},\mathsf z^{-3},\ldots,\mathsf z^{-(h-1)}\ \ \text{at infinity},\qquad \zeta\ \ \text{at zero},
\end{equation}
as possible violations of \eqref{eq:endpoints}. They are absent from the density \eqref{eq:densitydecay}, so their coefficients in the primitive are constant in $x$.

\subsection{Normalization by the integration constants}

The leading integrated string equations are hodograph relations for the Laurent coefficients of the partner,
\begin{equation}\label{eq:hodograph}
 [\mathsf z^{-k}]\,\pi\bigl(p(\mathsf z)\bigr)=\gamma_k-\delta_{k,h-1}\frac{\kappa x}{h},\quad k=1,3,\ldots,h-1,\qquad
 [\zeta]\,\pi\bigl(p(\zeta)\bigr)=\gamma_0,
\end{equation}
with $p=\mathsf z+O(\mathsf z^{-1})$ at infinity and $p=\rho_0\zeta+O(\zeta^3)$ at zero. Their quantum versions contain one constant for each forbidden mode in \eqref{eq:forbidden}. Shifting the constant attached to a mode changes its primitive coefficient by $-1/\kappa$ times the shift and leaves the other modes alone. Adjusting the constants genus by genus therefore removes every forbidden mode and produces an admissible solution with the original generator. This is what we call a normalized string.

To pass from strings to arbitrary curves we need the leading jets of normalized strings to fill an open set. Invertibility of the hodograph Jacobian controls only the values of the fields at a point. Their $x$-derivatives are supplied by the generator coefficients $\mathsf T_m$, as follows. Use coordinates $(u_1,\ldots,u_{n-1},d_0)$ with $d_0=\rho_0^2$, write $\boldsymbol{\mathsf h}$ for the vector of left sides of \eqref{eq:hodograph}, and form
\begin{equation}\label{eq:jetvector}
 \boldsymbol g=\Bigl(\frac1{W_p},\frac{p^2}{W_p},\ldots,\frac{p^{h-2}}{W_p},\frac{p^{-2}}{W_p}\Bigr)^t .
\end{equation}
For odd $m$ the variation of the hodograph vector with $\mathsf T_m$ is
\begin{equation}\label{eq:timecolumns}
 \partial_{\mathsf T_m}\boldsymbol{\mathsf h}=\mathsf M\,[\mathsf z^{-(m+h)}]\,\boldsymbol g,\qquad
 \mathsf M=\operatorname{diag}(\mathsf M_0,h\rho_0),\qquad
 (\mathsf M_0)_{ki}=-\frac hk[p^{2i-2}]\,\partial_p\bigl(W^{k/h}\bigr)_+,
\end{equation}
with $k=1,3,\ldots,h-1$. The ordinary block is triangular with diagonal $-h$, so $\mathsf M$ is invertible on the regular domain.

These columns span every finite jet order. Fix an order $N$ and a regular leading jet at which $W$ has a simple zero, and suppose that a linear functional annihilates the positive Taylor jets of all sufficiently high odd Laurent modes in \eqref{eq:timecolumns}. Absorbing $\mathsf M$ into an $x$-dependent basis $\mathsf b_i(p,x)$ of $\{p^{-2},1,p^2,\ldots,p^{h-2}\}$, the functional produces the rational function
\begin{equation}\label{eq:annihilator}
 \mathfrak a(p)=\sum_{i}\sum_{j=1}^{N}\mathsf l_{i,j}\,\cD^j\Bigl(\frac{\mathsf b_i(p,x)}{W_p}\Bigr)\Big|_{x=x_0},
\end{equation}
which is odd and of order $\mathsf z^{-1}$. Since all sufficiently high odd Laurent coefficients vanish, $\mathfrak a=\sum_k\mathsf c_kW^{-k/h}$ is a finite sum over odd $k$ near infinity. Continuing to a simple zero of $W$ and grouping the terms by $k$ modulo $h$, every group has a fractional exponent with nontrivial monodromy, because $h$ is even and $k$ is odd. A rational function has none, so every $\mathsf c_k$ vanishes and $\mathfrak a=0$. Now let $j$ be the highest derivative in \eqref{eq:annihilator}. Its highest pole at the ramification points has numerator proportional to $(2j-1)!!\,(p^3W_xT_p)^j\,p^3\sum_i\mathsf l_{i,j}\mathsf b_i$ modulo $T$. The first factor is invertible modulo $T$ by \eqref{eq:regular}, and the remaining odd polynomial has degree below $\deg T=2n$, so it vanishes. Independence of the $\mathsf b_i$ gives $\mathsf l_{i,j}=0$, and descending in $j$ shows that the functional vanishes. Finitely many $\mathsf T_m$ therefore span each finite jet order, while the $n$ integration constants supply the field values at $x_0$.

The open set is nonempty. Take $\mathsf T_1$ large, the ordinary coefficients of $W$ near zero and $d_0\neq0$. The leading hodograph Jacobian is then dominated by the invertible map to the first $n-1$ odd inverse coefficients of $p(\mathsf z)$ and to $\rho_0$, the critical points deform those of $p^{2n}=2d_0/h$, and $W_x$ is close to $\kappa/\mathsf T_1\neq0$. The analytic inverse function theorem gives an open image in finite-jet space. The compatibility residuals of the reconstruction are rational in finitely many jets and vanish on this open set, so their numerators are polynomial identities. The inversions used in the pole descent introduce denominators only where \eqref{eq:regular} fails, because the highest-pole multiplier is a nonzero integer times $p\,(p^2W_x)\,T_p$ in the quotient algebra by $T$, and each factor is invertible there. This completes the sketch of proposition~\ref{thm:regular}.

\subsection{The symmetric limit and the first Pfaffian direction}\label{app:symmetriclimit}

Proposition~\ref{thm:symmetriclimit} rests on the structure of the equations for the new fields. On a normalized string family, endpoint normalization turns the quantum hodograph relations into a linear system at each genus,
\begin{equation}\label{eq:limitmechanism}
 J\,\boldsymbol u_g=-\boldsymbol K_g,\qquad \boldsymbol u_g=(u_{1,g},\ldots,u_{n-1,g},\rho_g),
\end{equation}
where $J$ is the classical hodograph Jacobian of the $n$ leading equations \eqref{eq:hodograph} with respect to the $n$ fields, and $\boldsymbol K_g$ is a differential polynomial in lower-genus fields. The coefficients of $\boldsymbol K_g$ are regular at $\rho=0$. If $J$ is invertible at $\eta=\gamma_0=0$ and the constants are analytic in $\eta$, analyticity of $J^{-1}$ propagates from one genus to the next, and the reflection $\rho\to-\rho$, $\eta\to-\eta$ gives the stated parities. In the limit, the regular-domain uniqueness argument is replaced by its symmetric-sector version, in which the primitive may have a pole at $p=0$ because the origin has become a ramification point.

For the conformal strings $W=e_h$, $\pi=e_s$ with $h$ even and $s$ odd and coprime to $h$, the hypotheses can be checked in closed form. At $\rho=0$ the Jacobian is block diagonal, with origin entry $\pi_p(0)=s(-1)^{(s-1)/2}$. An ordinary variation in its kernel would have a fixed-energy variation $\delta\pi-\pi_p\delta W/W_p$ of order $p^{-h-1}$. Multiplying by $W_p$ gives a polynomial that decays at infinity and therefore vanishes, and evaluating at the $h-1$ roots $p_j=2\cos\theta_j$, $\theta_j=\pi j/h$, forces $\delta W=0$ because
\begin{equation}\label{eq:chebyshevcheck}
 \pi_p(p_j)=s\,\frac{\sin s\theta_j}{\sin\theta_j}\neq0,\qquad v_j=\frac\kappa{\pi_p(p_j)}\neq0 .
\end{equation}

The first Pfaffian generator is $\widehat B_1=-D^{-1}\rho$, and at the symmetric point it moves the origin constant at a universal rate,
\begin{equation}\label{eq:origintangent}
 \partial_{\widehat t_1}\gamma_0\big|_{\rho=0}=\frac\kappa2 .
\end{equation}
At $\rho=0$ the operator $L_D$ is differential and its symmetric ordering gives $L_D(1)=\h\,\partial_{\widehat t_1}\rho$. Differentiating $\gamma_0=\mathcal A(\rho)$ then gives $\h\,\partial_{\widehat t_1}\gamma_0=\mathcal Y(1)$ with $\mathcal Y=\mathcal AL_D$, and the pencil form of the string equation becomes $D\mathcal Y+\mathcal Y^*D=\kappa\h D$. In Weyl symbols this forces the odd and even parts of $\operatorname{symb}(\mathcal Y)$ to cancel pairwise in $\mathcal Y(1)$, leaving exactly $\kappa\h/2$ at every order in $\h$. Equation \eqref{eq:origintangent} fixes the normalization of the Pfaffian coupling used in appendix~\ref{app:worldsheet}.
\section{The genus-one derivation}\label{app:genusone}

This appendix derives proposition~\ref{prop:moving-genus-one}, given proposition~\ref{prop:quantum-tau}, together with the top-field part of its scalar analogue \eqref{eq:FoneA}. Subscripts on $W$ denote partial derivatives, each holding the other variable fixed. On a sheet, $\partial_\lambda=W_p^{-1}\partial_p$ commutes with $\cD=\partial_x-W_x\partial_\lambda$. For a root $a(x)$ of $T=p^3W_p$ put $v_a=W_x(a)$ and $\ell_a=\dd\log v_a/\dd x$. Differentiating $W_p(a)=0$ gives
\begin{equation}\label{eq:aprime}
 a'=-\frac{W_{px}(a)}{W_{pp}(a)},\qquad v_a'=W_{xx}(a)-\frac{W_{px}(a)^2}{W_{pp}(a)} .
\end{equation}

\paragraph{The candidate.} Equation \eqref{eq:Qstar} defines $\Qhat_*=\cB^{(0)}(\lambda)F_*$. Both $\Upsilon$ and $\partial_p\log T$, which equals $W_{pp}/W_p+3/p$, are odd, of order $O(p)$ at zero and $O(p^{-1})$ at infinity, and $\cD$ preserves these bounds. Applying $\partial_\lambda$ therefore gives $\Qhat_*=O(p^3)$ at zero and $O(p^{-h-1})$ at infinity. Each derivative raises the order of a pole at a simple ramification point by at most two, so $\Qhat_*$ has poles of order at most five. These are the endpoint conditions of section~\ref{sec:endpoints} at genus one.

\paragraph{The genus-one density.} Set
\begin{equation}\label{eq:GammaXi}
 \Gamma=W_{pp}W_{xx}-W_{px}^2,\qquad
 \Xi=W_{pp}W_x^2-2W_{px}W_pW_x+W_{xx}W_p^2,
\end{equation}
and define
\begin{equation}\label{eq:frake}
 \mathfrak e=\frac18\Bigl(\partial_p^2+\frac2p\,\partial_p+\frac1{p^2}\Bigr)W_{xx}-\frac{(\partial_x\rho_0)^2}{p^4}
 =\frac{O_D}p+\frac{W_{xx}}{8p^2},
\end{equation}
where the second equality uses \eqref{eq:aord}. Substituting $\mathsf A_0=p(W-\lambda)$ and $\mathsf A_2=p\Phi_1+O_D$ in \eqref{eq:G2} and collecting powers of $W-\lambda$ gives
\begin{equation}\label{eq:pG2}
 pG_2=-\frac{\Phi_1+\mathfrak e+W_{xx}/(8p^2)}{(W-\lambda)^2}+\frac{\Gamma+W_x^2/p^2}{4(W-\lambda)^3}-\frac{\Xi}{4(W-\lambda)^4}.
\end{equation}
The residue rule \eqref{eq:residuerule} then gives $24\Rhat_2=\partial_\lambda\mathfrak k$, with
\begin{equation}\label{eq:frakk}
 \mathfrak k=-\frac{24(\Phi_1+\mathfrak e)}{W_p}+3\,\partial_\lambda\frac\Gamma{W_p}-\partial_\lambda^2\frac\Xi{W_p}+3\,\partial_\lambda\frac{W_x^2}{p^2W_p}-\frac{3W_{xx}}{p^2W_p}.
\end{equation}

\paragraph{Two identities.} Substituting $\cD=\partial_x-(W_x/W_p)\partial_p$, one checks
\begin{align}\label{eq:twoidentities}
 W_p\Bigl[\cD^2\frac{W_{pp}}{W_p}+3\,\partial_\lambda\frac\Gamma{W_p}-\partial_\lambda^2\frac\Xi{W_p}\Bigr]&=\partial_p\frac\Gamma{W_p},\\
 \cD^2\frac1p+\partial_\lambda\frac{W_x^2}{p^2W_p}&=\frac{W_{xx}}{p^2W_p}.\nonumber
\end{align}
Both hold for an arbitrary function $W(p,x)$. By \eqref{eq:Qstar}, $\cD \Qhat_*=\Rhat_2$ holds once $\mathfrak k=-\cD\Upsilon-\cD^2\partial_p\log T$, and the two identities turn this condition into the field correction \eqref{eq:Phi1},
\begin{equation}\label{eq:Phi1app}
 24\Phi_1=\partial_p\frac\Gamma{W_p}+W_p\,\cD\Upsilon-24\,\mathfrak e .
\end{equation}
Nothing in this step uses the specific form of $\Upsilon$. The equation $\cD Q=\Rhat_2$ is solved by $24Q=-\partial_\lambda(\Upsilon+\cD\partial_p\log T)$ and \eqref{eq:Phi1app} for every odd $\Upsilon$.

\paragraph{Admissibility.} It remains to see which $\Upsilon$ gives a field correction of the form \eqref{eq:Phi}. Near a ramification point, $\Gamma/W_p=v_a'/(p-a)+O(1)$. The singular part of $W_p\cD\Upsilon$ is
\begin{equation}
 \frac{W_p\,\ell_a'}{p-a}+\frac{\ell_a\,(W_pa'+W_x)}{(p-a)^2}.
\end{equation}
The first term is regular, and $W_pa'+W_x=v_a+O((p-a)^2)$ because $W_{pp}(a)a'+W_{px}(a)=0$. The double pole therefore has coefficient $\ell_av_a=v_a'$ and cancels the one in $\partial_p(\Gamma/W_p)$. At the origin put $d_0=\rho_0^2$. The leading terms are
\begin{equation}
 \frac\Gamma{W_p}=\Bigl(-3\,\partial_x^2d_0+\frac{2(\partial_xd_0)^2}{d_0}\Bigr)p^{-3}+O(p^{-1}),\qquad
 24\,\mathfrak e=\bigl(9\,\partial_x^2d_0-24(\partial_x\rho_0)^2\bigr)p^{-4}+O(p^{-2}),
\end{equation}
and $(\partial_xd_0)^2/d_0=4(\partial_x\rho_0)^2$, so the fourth-order poles cancel. The term $W_p\cD\Upsilon$ is $O(p^{-2})$ there. At infinity, $\mathfrak e$ and $\partial_p(\Gamma/W_p)$ are $O(p^{h-4})$. The coefficient of $1/p$ in $\Upsilon$ is $\sum_a\ell_a$, and since $\cD$ preserves parity and $(W_x/W_p)\partial_p\Upsilon=O(p^{-3})$, one finds $\cD\Upsilon=(\sum_a\ell_a')/p+O(p^{-3})$. Hence
\begin{equation}
 \Phi_1=\frac h{24}\Bigl(\sum_a\ell_a'\Bigr)p^{h-2}+O(p^{h-4}).
\end{equation}
The even function \eqref{eq:Phi1app} therefore lies in the span of $p^{-2},1,p^2,\ldots,p^{h-2}$, and its coefficient of $p^{-2}$ fixes $2\rho_0\rho_1$. The pair $(\Phi_1,\Qhat_*)$ solves the genus-one reconstruction problem, and uniqueness identifies it with the reconstruction. Its top coefficient is
\begin{equation}
 u_{n-1,1}=\frac h{24}\sum_a\frac{\dd^2}{\dd x^2}\log v_a=2h\,\partial_x^2F_* .
\end{equation}
This determines $F^D_1$ only up to terms annihilated by $\partial_x^2$. On a normalized string, $\Qhat_*=\cB^{(0)}F_*$ by construction and $\Qhat_*=\Qhat_1$ by uniqueness, while proposition~\ref{prop:quantum-tau} identifies $\Qhat_1$ with $\cB F^D_1$. Comparing the Laurent coefficients fixes every ordinary first derivative of $F^D_1$, which proves proposition~\ref{prop:moving-genus-one}.

\paragraph{The scalar case.} For the scalar operator of Part I, $Q_A=D^h+\sum_j\tfrac12\{u^A_j,D^j\}$, the symbol is $W-\lambda+\h^2(\Phi_1+O_A)$, with ordering term $O_A=\sum_jj(j-1)u^A_{j,0}{}''p^{j-2}/8$, and the density carries no momentum insertion. The same steps go through with $\mathsf A_0=W-\lambda$. One finds $G_2=-(\Phi_1+O_A)/(W-\lambda)^2+\Gamma/(4(W-\lambda)^3)-\Xi/(4(W-\lambda)^4)$, and only the first identity of \eqref{eq:twoidentities} is needed. The primitive is $24Q=-\partial_\lambda(\Upsilon+\cD\,\partial_p\log W_p)$ with the sum in $\Upsilon$ over the $h-1$ roots of $W_p$, and the field correction is $24\Phi_1=\partial_p(\Gamma/W_p)+W_p\cD\Upsilon-24O_A$, a polynomial of degree $h-2$ with top coefficient $(h/24)\sum_a\ell_a'$. With the scalar tau normalization $u^A_{h-2,g}=h\,\partial_x^2F^A_g$ this fixes $F^A_1$ up to terms annihilated by $\partial_x^2$. The first derivatives along the ordinary times follow from the scalar counterpart of proposition~\ref{prop:quantum-tau}, whose argument uses only the bilinear identity of the reduced KP hierarchy in place of the BKP one, the transport of the string and uniqueness \cite{BertolaDubrovinYang2021}. For $h=2$ it is the loop-operator identity of ref.~\cite{Johnson2026Universal}. This gives \eqref{eq:FoneA} up to a constant independent of the ordinary times. We have not written out the scalar argument in full, and the A-series statements of appendix~\ref{app:worldsheet} rest on it.

\paragraph{Checks.} The algebraic steps \eqref{eq:pG2}--\eqref{eq:Phi1app} were verified symbolically for a generic function $W(p,x)$ and a generic odd $\Upsilon$, in both the pencil and the scalar case, and perturbed coefficients fail. The admissibility statements were checked on random rational jets for $D_3$, $D_4$ and $D_5$ and for scalar operators with $h=3,\ldots,6$ at sixty digits. On the regular $D_3$ string of appendix~\ref{app:regular}, \eqref{eq:Qstar} reproduces the primitive obtained there from the corrected trace.
\section{Boundary insertions}\label{app:boundaries}

This appendix fills in section~\ref{sec:boundaries}. We first explain why a normalized string stays normalized under the ordinary flows, then sketch the matching with the tau function, and finally derive the cylinder and the three-boundary formula and state the loop equations.

\subsection{Transport of the string and the tau function}

A string background is not invariant under the flows, because the generator $S$ in \eqref{eq:string} depends on the times through the fields. Using $\sum_k\mathsf T_k\partial_{t_k}\cL=\kappa$, one finds
\begin{equation}\label{eq:transportobstruction}
 \partial_{t_j}S-\frac{[B_j,S]}\h=\sum_k(\partial_{t_j}\mathsf T_k)B_k+\frac{\kappa j}{h}\bigl(P^{j-h}\bigr)_+ .
\end{equation}
The string is therefore transported covariantly, $\partial_{t_j}S=[B_j,S]/\h$, when the generator coefficients evolve as
\begin{equation}\label{eq:transporttimes}
 \partial_{t_j}\mathsf T_k=-\frac{\kappa j}{h}\,\delta_{k,j-h}.
\end{equation}
For $j<h$ nothing moves, while a higher flow shifts the coefficient of $B_{j-h}$. The Jacobi identity then preserves $[S,\cL]=\kappa\h$. Following a common wavefunction for the spatial, spectral and flow equations, one reads off from the quantum hodographs the endpoint constants $\mathcal H^\infty_k$ of the modes $\mathsf z^{-k}$ and $\mathcal H^0$ of the mode $\zeta$ in \eqref{eq:forbidden}, and finds
\begin{equation}\label{eq:endpointflow}
 \partial_{t_j}\mathcal H^\infty_k=-\frac{\kappa j}h\,\delta_{k,h-j},\qquad \partial_{t_j}\mathcal H^0=0 .
\end{equation}
The only change is the classical shift of one leading constant. Every positive-genus endpoint constant is conserved, so a normalized string remains normalized along the flows. Combined with uniqueness, this shows that inserting a boundary commutes with reconstruction. The quantum insertion acting on reconstructed fields equals the dispersionless insertion acting on the leading jets, followed by reconstruction.

The rest of this subsection sketches the proof of proposition~\ref{prop:quantum-tau}. We first build a primitive from the tau function and then use the string to fix its normalization. The Pfaffian times are held fixed and suppressed. Write $\mathfrak t(\mathsf z)=\sum_{j>0\ {\rm odd}}t_j\mathsf z^j$, and let $[\mathsf z^{-1}]$ be the shift of the ordinary times with components $1/(j\mathsf z^j)$. The ordinary Baker--Akhiezer wavefunction of the D hierarchy is \cite{LiuWuZhang2009}
\begin{equation}\label{eq:BAwave}
 \psi(\mathsf z)=e^{\mathfrak t(\mathsf z)/\h}\,\frac{\tau_D(\boldsymbol t-2\h[\mathsf z^{-1}])}{\tau_D(\boldsymbol t)},\qquad P\psi(\mathsf z)=\mathsf z\,\psi(\mathsf z).
\end{equation}
Take two energies $\lambda_i=\mathsf z_i^h$ and put $\psi_1=\psi(\mathsf z_1)$ and $\psi_2=\psi(-\mathsf z_2)$. Since $h$ is even, $\psi_2$ solves the pencil at energy $\lambda_2$. The pencil has a Lagrange concomitant $\mathfrak C_L(\psi_2,\psi_1)$, the analogue of a Wronskian. It is built from the differential part of $L_D$ in the usual way, together with the product $\chi_1\chi_2$ of the auxiliary components $\chi_i=D^{-1}(\rho\psi_i)$, which accounts for $\rho D^{-1}\rho$. Since $L_D$ is antisymmetric, $D\mathfrak C_L=\psi_2L_D\psi_1+\psi_1L_D\psi_2$. Subtracting the averaged energy gives the two-energy pairing
\begin{equation}\label{eq:twoenergy}
 \mathfrak C(\mathsf z_1,\mathsf z_2)=\mathfrak C_L(\psi_2,\psi_1)-\frac{\lambda_1+\lambda_2}2\,\psi_1\psi_2,\qquad
 D\mathfrak C=\frac{\lambda_1-\lambda_2}2\bigl(\psi_2D\psi_1-\psi_1D\psi_2\bigr).
\end{equation}
The squared-eigenfunction identity of the BKP hierarchy \cite{ChengHeHu2010} gives a tau expression with the same derivative. The integration constant can be fixed within the formal wave expansion, and one finds
\begin{align}\label{eq:twoenergytau}
 \mathfrak C(\mathsf z_1,\mathsf z_2)&=\frac{(\mathsf z_1^h-\mathsf z_2^h)(\mathsf z_1+\mathsf z_2)}{2(\mathsf z_1-\mathsf z_2)}\,e^{[\mathfrak t(\mathsf z_1)-\mathfrak t(\mathsf z_2)]/\h}\,\frac{\tau_D(\boldsymbol t+2\h[\mathsf z_2^{-1}]-2\h[\mathsf z_1^{-1}])}{\tau_D(\boldsymbol t)},\nonumber\\
 \mathfrak C(\mathsf z,\mathsf z)&=h\mathsf z^h .
\end{align}
Differentiating in the first energy at coincidence defines the bilinear primitive
\begin{equation}\label{eq:bilinearprimitive}
 \mathcal Q(\lambda)=\h\,\frac{\partial_{\lambda_1}\mathfrak C(\mathsf z_1,\mathsf z_2)}{\mathfrak C(\mathsf z_1,\mathsf z_2)}\bigg|_{\mathsf z_1=\mathsf z_2=\mathsf z},
\end{equation}
which can be evaluated in two ways. In the basis \eqref{eq:localY} the pairing reads
\begin{equation}\label{eq:pairinglocal}
 \mathfrak C(\mathsf z_1,\mathsf z_2)=Y(-\mathsf z_2)^t\,\mathsf J\bigl(\tfrac12(\lambda_1+\lambda_2)\bigr)\,Y(\mathsf z_1),
\end{equation}
where the matrix $\mathsf J$ depends on the energy only through the subtracted term, so that $\partial_{\lambda_1}\mathsf J$ is minus one half of the matrix unit at position $(1,1)$. At coincidence the row $\widetilde Y=Y(-\mathsf z)^t\mathsf J(\lambda)$ is a dual solution, $\h\partial_x\widetilde Y=-\widetilde Y\mathsf L$, and $M=Y(\mathsf z)\widetilde Y/\mathfrak C(\mathsf z,\mathsf z)$ is the normalized projector onto the ordinary sheet, with $M_{1,h+1}=\psi(\mathsf z)\psi(-\mathsf z)/\mathfrak C(\mathsf z,\mathsf z)$. Hence
\begin{equation}\label{eq:bilinearprojector}
 \mathcal Q=\h\,\frac{\widetilde Y\partial_\lambda Y}{\widetilde YY}-\frac\h2M_{1,h+1},\qquad
 \partial_x\mathcal Q=M_{2,h+1}-\frac\h2\partial_xM_{1,h+1}=\Rhat .
\end{equation}
The $x$-derivative of the first term is $\tr\bigl((\partial_\lambda\mathsf L)M\bigr)=M_{2,h+1}$, and the last equality is \eqref{eq:projectordensity}. The ordering subtraction in the corrected trace primitive \eqref{eq:correctedtrace} is therefore the energy derivative of the averaged energy in \eqref{eq:twoenergy}. On the other side, the logarithmic $\mathsf z_1$-derivative of \eqref{eq:twoenergytau} at coincidence receives $h/(2\mathsf z)$ from the prefactor and $2\h\sum_{j>0\ {\rm odd}}\mathsf z^{-j-1}\partial_{t_j}\log\tau_D$ from the shifted tau function. Dividing by $\dd\lambda/\dd\mathsf z=h\mathsf z^{h-1}$ gives
\begin{equation}\label{eq:bilineartau}
 \mathcal Q=\frac{\partial_{\mathsf z}\mathfrak t(\mathsf z)}{h\mathsf z^{h-1}}+\frac\h{2\lambda}+\h^2\cB(\lambda)\log\tau_D ,
\end{equation}
with the loop operator \eqref{eq:loopoperator}. Its coefficient $2/h$ is the factor $2\h$ of the tau shift divided by this Jacobian. The first term is classical and the second is odd in $\h$. Differentiating in $x=t_1$ and using \eqref{eq:tau} gives the exact quantum density identity
\begin{equation}\label{eq:quantumdensity}
 \Rhat(\lambda)-\frac1h\mathsf z^{1-h}=\frac1h\sum_{j>0\ {\rm odd}}\mathsf z^{-h-j}\res_DP^j
\end{equation}
on any even solution of the D hierarchy.

The density identity fixes $\mathcal Q$ only up to a function of the energy, and that function could depend on the other times. On a string family it does not. The wavefunction \eqref{eq:BAwave} need not obey the spectral equation $\kappa\h\partial_\lambda\psi=-S\psi$ of appendix~\ref{app:reconstruction}, and its defect is
\begin{equation}\label{eq:defect}
 \mathcal E(\mathsf z;\h)=\frac{(\h\partial_\lambda+S/\kappa)\psi(\mathsf z)}{\psi(\mathsf z)} .
\end{equation}
Differentiating the monic root $P=\cL^{1/h}$ along the string commutator gives $[S,P]=\kappa\h P^{1-h}/h$, so the numerator $(\h\partial_\lambda+S/\kappa)\psi$ is annihilated by $P-\mathsf z$. Writing $P=\mathcal U D\mathcal U^{-1}$ and $\psi=\mathcal U e^{\mathfrak t(\mathsf z)/\h}$, with the dressing operator $\mathcal U$, this numerator is $\mathcal E\,\psi$ with $\partial_x\mathcal E=0$. The transport \eqref{eq:transporttimes} gives
\begin{equation}\label{eq:defecttransport}
 \bigl[\h\partial_{t_j}-B_j,\ \h\partial_\lambda+S/\kappa\bigr]=\frac\h\kappa\partial_{t_j}S-\frac1\kappa[B_j,S]=0,
\end{equation}
and applying it to $\mathcal E\,\psi$ gives $\partial_{t_j}\mathcal E=0$. The defect is therefore independent of $x$ and of every ordinary time. Comparing with the spectral connection of the string gives $\mathcal Q=\Qhat+\mathcal E$, and at positive genus \eqref{eq:bilineartau} then gives
\begin{equation}\label{eq:lineargauge}
 \cB F^D_g-\Qhat_g=\mathcal E_g(\mathsf z)=\frac2h\sum_{j>0\ {\rm odd}}\alpha_{g,j}\,\mathsf z^{-h-j},\qquad
 F^D_g\longmapsto F^D_g-\sum_{j>0\ {\rm odd}}\alpha_{g,j}t_j,
\end{equation}
with coefficients $\alpha_{g,j}$ that are constant along the family. The displayed shift is a linear tau gauge. It preserves the bilinear identity, all fields and all connected derivatives of order two or more, and it removes $\mathcal E_g$ coefficient by coefficient. This gives proposition~\ref{prop:quantum-tau} as a formal series at large energy, and its positive-genus coefficients continue to the rational functions produced by reconstruction.

\subsection{Genus zero and the loop equations}

At genus zero the insertion of a boundary at energy $\lambda_0=W(p_0)$ moves the disk by $\cD$ of $\Omega_{\lambda_0}$, with $\Omega$ from \eqref{eq:dispersionless}. With the disk reference whose variation is $\partial_p\Omega_{\lambda_0}/W_p$, this gives
\begin{equation}\label{eq:cylinderderivation}
 \cB^{(0)}(\lambda_0)\,\omega_{0,1}(p)\,\dd\lambda_0=\partial_p\Bigl(\frac{2p}{p_0^2-p^2}\Bigr)\dd p_0\,\dd p=B_D(p,p_0),
\end{equation}
which is \eqref{eq:cylinder}. On a normalized string the reference is supplied by the classical time dependence of the wavefunction. Differentiating the explicit time polynomial $\mathfrak t$ introduced above, with $\mathsf z_i=\lambda_i^{1/h}$, and inserting once more gives
\begin{equation}\label{eq:contact}
 \cB(\lambda_2)\frac{\partial_{\mathsf z_1}\mathfrak t(\mathsf z_1)}{h\,\mathsf z_1^{h-1}}=\frac2{h^2}\sum_{j>0\ {\rm odd}}j\,\mathsf z_1^{j-h}\mathsf z_2^{-j-h}=\frac{(\mathsf z_1-\mathsf z_2)^{-2}+(\mathsf z_1+\mathsf z_2)^{-2}}{h^2\mathsf z_1^{h-1}\mathsf z_2^{h-1}},
\end{equation}
a term independent of the fields that completes the raw tau Hessian into the paired kernel. Any further insertion annihilates it.

Write $b(p,q)=(p-q)^{-2}+(p+q)^{-2}$, $\mathcal V(p)=W_x(p)/W_p(p)$ and $\mathsf F(p_0,p)=2p/(p_0^2-p^2)$, and let
\begin{equation}\label{eq:threeH}
 \Theta(p)=\partial_p\mathsf F(p_0,p)\,\mathcal V(p)+\partial_{p_0}\mathsf F(p_0,p)\,\mathcal V(p_0)+\mathsf F(p_0,p)\,\mathcal V'(p_0).
\end{equation}
Differentiating $\Omega_{\lambda_0}=\mathsf F/W_p(p_0)$ at fixed insertion energy gives
\begin{equation}
 \cB^{(0)}W(p)=\frac{W_p(p)\,\Theta(p)}{W_p(p_0)} .
\end{equation}
A second boundary point $p_1$ therefore moves by $\delta p_1=-\Theta(p_1)/W_p(p_0)$, so its line element changes by $\delta(\dd p_1)/\dd p_1=-\Theta'(p_1)/W_p(p_0)$. Applying these two variations to the cylinder $B_D(p_1,p_2)$ and simplifying gives
\begin{equation}\label{eq:threerational}
 \frac{\omega_{0,3}(p_0,p_1,p_2)}{\dd p_0\dd p_1\dd p_2}=-\sum_{i=0}^2\Bigl[\mathcal V'(p_i)\prod_{j\neq i}b(p_i,p_j)+\mathcal V(p_i)\,\partial_{p_i}\prod_{j\neq i}b(p_i,p_j)\Bigr].
\end{equation}
Now consider the differential $\mathcal V(a)\prod_ib(a,p_i)\,\dd a$. On the regular domain it is regular at zero and has no residue at infinity. Its remaining poles are the simple ramification points and the marked pairs $\pm p_i$. The residue at a ramification point is $W_x(a)\prod_ib(a,p_i)/W_{pp}(a)$, and the residues at $p_i$ and $-p_i$ add up to twice the $i$th bracket of \eqref{eq:threerational}. The residue theorem gives \eqref{eq:threeboundaries}.

Choose a disk around one representative $a$ of each opposite pair of ramification points, with the local deck transformation $\sigma_a$ and the disk $y=-\pi/\kappa$. The mirror recursion is the residue recursion
\begin{align}\label{eq:mirror}
 \omega^{\rm mir}_{g,k+1}(p_0,\boldsymbol p)=\sum_a\Res_{p\to a}\mathsf K_a(p_0,p)\Bigl[&\omega^{\rm mir}_{g-1,k+2}(p,\sigma_a(p),\boldsymbol p)\\
 &+\sideset{}{'}\sum_{\substack{g_1+g_2=g\\ I\sqcup J=\{1,\ldots,k\}}}\omega^{\rm mir}_{g_1,|I|+1}(p,\boldsymbol p_I)\,\omega^{\rm mir}_{g_2,|J|+1}(\sigma_a(p),\boldsymbol p_J)\Bigr],\nonumber
\end{align}
with kernel
\begin{equation}\label{eq:mirrorkernel}
 \mathsf K_a(p_0,p)=\frac{\int_{\sigma_a(p)}^pB_D(p_0,\cdot)}{2\bigl(y(p)-y(\sigma_a(p))\bigr)\,\dd W(p)},
\end{equation}
initial data $\omega_{0,1}=y\,\dd W$ and $\omega_{0,2}=B_D$, and the prime omitting terms with $\omega_{0,1}$. It is the Eynard--Orantin recursion \cite{EynardOrantin2007} with the Bergman kernel replaced by the paired kernel.

On regular normalized string families the loop differentials \eqref{eq:connected} have three of the properties that characterize this recursion. They are symmetric rational differentials whose poles lie only at the ramification points, they are regular at zero and infinity and odd under $p_i\to-p_i$, and in a local coordinate with $W=W(a)+\xi^2/2$ they satisfy the linear loop equation
\begin{equation}\label{eq:linearloop}
 \omega_{g,k}(p,\boldsymbol p)+\sigma_a^*\omega_{g,k}(p,\boldsymbol p)=O(\xi)\,\dd\xi .
\end{equation}
\paragraph{Sketch of proposition~\ref{prop:looplinear}.} Rationality and the pole locations propagate through every insertion. The field variation generated by $\Omega_{\lambda_0}$ stays in the finite D field space, its apparent poles at $p=\pm p_0$ cancel, and multiplication by $\dd\lambda_0$ makes it regular at $p_0=0$ and $p_0=\infty$, because $\Omega_{\lambda_0}\dd\lambda_0=2p\,\dd p_0/(p_0^2-p^2)$. At an existing marked point the fixed-energy variation of a one-form is its coefficient variation minus a Lie derivative along $(\delta W/W_p)\partial_p$, which is $O(p)$ at zero and $O(p^{-1})$ at infinity and so preserves endpoint regularity.

The linear loop equation is a statement about the local exchange $\sigma_a$, which swaps the two sheets meeting at $a$ and fixes the other sheets, including the pair meeting at $-a$. In the localized system \eqref{eq:localY} two leading eigenvalues meet at $a$, and their two-dimensional spectral subspace stays separated from all the others. Its projector $\Pi_a=M(p)+M(\sigma_a(p))$, with $M(p)$ the projector of appendix~\ref{app:reconstruction} onto the sheet through $p$, is holomorphic in $\lambda$ near $W(a)$, coefficientwise in $\h$. One builds it without separating the two eigenvalues inside the cluster. Start from the analytic spectral projector of the leading matrix onto the cluster and solve $\Pi_a^2=\Pi_a$ and $\h\partial_x\Pi_a=[\mathsf L,\Pi_a]$ order by order. Idempotence fixes the diagonal blocks, and Sylvester equations between the cluster and its complement fix the off-diagonal blocks. Their spectra are disjoint, so these equations have analytic inverses even at the ramification point. The corrected trace \eqref{eq:correctedtrace} then gives the holomorphic sum
\begin{equation}\label{eq:clustertrace}
 \Qhat(p)+\Qhat(\sigma_a(p))=\tr(\mathsf B\,\Pi_a)-\frac\h2(\Pi_a)_{1,h+1},
\end{equation}
and multiplying by $\dd\lambda=\xi\,\dd\xi$ proves \eqref{eq:linearloop} for one boundary. The construction is analytic in the energy and in the ordinary deformation parameters, so further insertions preserve the property. At genus zero the same statement follows from \eqref{eq:threeboundaries}, whose only singular term at the selected branch point is a double pole with no residue in $\xi$. The global reflection $p\to-p$ plays a different role. It relates the clusters at $a$ and $-a$ and gives the oddness of each $\omega_{g,k}$.

Rationality and reflection parity then give the projection property with the paired kernel. Adding the quadratic loop equation, which states that the combination
\begin{multline}\label{eq:quadraticloop}
 \omega_{g-1,k+2}(p,\sigma_a(p),\boldsymbol p)+\sideset{}{'}\sum\omega_{g_1,|I|+1}(p,\boldsymbol p_I)\,\omega_{g_2,|J|+1}(\sigma_a(p),\boldsymbol p_J)\\
 +\omega_{0,1}(p)\,\sigma_a^*\omega_{g,k+1}+\sigma_a^*\omega_{0,1}(p)\,\omega_{g,k+1}
\end{multline}
is $O(\xi^2)(\dd\xi)^2$ at every ramification point, would identify the loop differentials with \eqref{eq:mirror} at every stable order, following the abstract loop-equation argument of ref.~\cite{BorotEynardOrantin2015} with the paired projection of the odd recursion \cite{AlexandrovShadrin2023}. The loop differentials agree with \eqref{eq:mirror} in every case we have computed. These cases are the three-boundary amplitude on random regular $D_3$ and $D_4$ jets, the four-boundary amplitude at genus zero, the Potts one-holed torus of section~\ref{sec:pottsg1}, and the genus-one two-boundary amplitude on a regular $D_3$ curve to twenty-six digits.
\section{A regular example at low rank}\label{app:regular}

Before the symmetric limit takes over, we run the whole recipe on a curve where the Pfaffian field is alive. The smallest case is $D_3$, which is isomorphic to $A_3$. It therefore also gives a precise dictionary between the pencil and a scalar fourth-order operator.

\paragraph{The string and its fields.}

For $n=3$ the pencil \eqref{eq:intro-pencil} reads
\begin{equation}\label{eq:D3pencil}
 L_D=D^5+\tfrac12\{u_2,D^3\}+\tfrac12\{u_1,D\}+\rho D^{-1}\rho,\qquad L_D\psi=\lambda D\psi,
\end{equation}
so $W=p^4+u_2p^2+u_1+\rho^2/p^2$. The third ordinary flow is generated by $S=(\cL^{3/4})_+=D^3+3u_2D/4$, and the string condition $[S,\cL]=\kappa\h$ integrates to three equations,
\begin{align}\label{eq:D3integrals}
 u_1&=\mathfrak c_1+\frac{u_2^2}8-\frac56\h^2u_2'',\nonumber\\
 \frac34u_2\rho+\h^2\rho''&=\gamma_0,\\
 \frac{u_2^3}{16}+3\rho^2+\h^2\Bigl(\frac38u_2u_2''+\frac3{16}(u_2')^2\Bigr)+\frac16\h^4u_2^{(4)}&=\kappa x+\mathfrak c_0,\nonumber
\end{align}
with integration constants $\mathfrak c_0$, $\mathfrak c_1$ and $\gamma_0$. The second line is the origin equation \eqref{eq:origin}. We take $\kappa=1$ and leading values $(u_{1,0},u_{2,0},\rho_0)=(1/2,2,1)$ at $x=0$, which fix $\mathfrak c_1=0$, $\gamma_0=3/2$ and $\mathfrak c_0=7/2$. The genus-zero solution is parameterized by $\mathsf t=u_{2,0}$,
\begin{equation}\label{eq:D3leading}
 u_{2,0}=\mathsf t,\qquad \rho_0=\frac2{\mathsf t},\qquad u_{1,0}=\frac{\mathsf t^2}8,\qquad x=\frac{\mathsf t^3}{16}+\frac{12}{\mathsf t^2}-\frac72 .
\end{equation}
Expanding \eqref{eq:D3integrals} in $\h$ gives a two-by-two linear system for $(u_{2,g},\rho_g)$ at each genus, with $u_{1,g}$ following from the first line. The moving genus-one fields are
\begin{align}\label{eq:D3g1}
 u_{2,1}(\mathsf t)&=\frac{256\mathsf t^4(\mathsf t^5+512)(9\mathsf t^5+128)}{27(\mathsf t^5-128)^4},\qquad
 \rho_1(\mathsf t)=-\frac{12800\mathsf t^7(\mathsf t^5+128)}{27(\mathsf t^5-128)^4},\nonumber\\
 u_{1,1}(\mathsf t)&=\frac{64\mathsf t^5(29\mathsf t^{10}+6016\mathsf t^5-425984)}{27(\mathsf t^5-128)^4},
\end{align}
and at the reference point $\mathsf t=2$ the fields through genus two are listed in table~\ref{tab:D3fields}.

\begin{table}[t]
\centering
\begin{tabular}{@{}cccc@{}}
\toprule
& $g=0$ & $g=1$ & $g=2$\\
\midrule
$u_{1,g}$ & $1/2$ & $-398/2187$ & $-32463374/14348907$\\
$u_{2,g}$ & $2$ & $884/2187$ & $83442200/14348907$\\
$\rho_g$ & $1$ & $-250/2187$ & $-24253600/14348907$\\
\bottomrule
\end{tabular}
\caption{Fields of the regular $D_3$ string \eqref{eq:D3integrals} at $x=0$. At genus one they agree with the field correction \eqref{eq:D3Phi} of the pencil reconstruction, whose $p^{-2}$ coefficient is $2\rho_0\rho_1$.}\label{tab:D3fields}
\end{table}

The velocity law can be seen directly on this orbit. The squares $a^2$ of the ramification points solve $2a^6+\mathsf ta^4-4/\mathsf t^2=0$, and at each of them the string equation gives $v_a=(3a^2+3\mathsf t/4)^{-1}$. Evaluating the cubic at $a^2=-\mathsf t/4$ gives the product over the three pairs,
\begin{equation}\label{eq:D3velocities}
 \prod_{a/\pm}v_a=-\frac{64\,\mathsf t^2}{27(\mathsf t^5-128)} .
\end{equation}
Taking $8\partial_x^2$ of one twenty-fourth of its logarithm, with $\partial_x=(\dd x/\dd \mathsf t)^{-1}\partial_{\mathsf t}$, reproduces $u_{2,1}(\mathsf t)$ of \eqref{eq:D3g1} on the whole orbit, as required by \eqref{eq:genusone} with $h=4$.

\paragraph{The scalar partner.}

The scalar operator $Q_A=D^4+\{u^A_2,D^2\}/2+\{u^A_1,D\}/2+u^A_0$ has a four-dimensional companion system, and its exterior square is six-dimensional. For two solutions $f_1,f_2$ put $X_{ij}=D^if_1D^jf_2-D^jf_1D^if_2$, $\psi=X_{01}$ and $\chi=X_{12}-X_{03}-u^A_2\psi$. Then $D\chi=u^A_1\psi$, and $\psi$ obeys a fifth-order equation which, after restoring the symmetric ordering, is the pencil \eqref{eq:D3pencil} with
\begin{equation}\label{eq:exactmap}
 u_2=2u^A_2,\qquad u_1=(u^A_2)^2-4u^A_0-2\h^2u^A_2{}'',\qquad \rho=iu^A_1,\qquad \lambda_D=-4\lambda_A .
\end{equation}
The induced generator is $S_A^{\wedge2}=-S_D/2$, so $\kappa_D=8\kappa_A$, and our background corresponds to $\kappa_A=1/8$ with
\begin{equation}
 (u^A_{2,0},u^A_{1,0},u^A_{0,0})=\Bigl(\frac{\mathsf t}2,-\frac{2i}{\mathsf t},\frac{\mathsf t^2}{32}\Bigr),\qquad E_A(k)=k^4+k^2-ik+\frac18
\end{equation}
at $x=0$. The classical momenta are related by $p=k_i+k_j$ for two scalar sheets at a common energy. The relation $u_{2,g}=2u^A_{2,g}$, together with the two tau normalizations, gives $F^D_g=F^A_g$, as it must for isomorphic theories. Up to a factor of $i$, the Pfaffian field of the pencil is the coefficient $u^A_1$ of the first-order term of the scalar operator.

\paragraph{Both one-holed tori.}

At $x=0$ the energy is $W=p^4+2p^2+1/2+p^{-2}$, and the polynomial \eqref{eq:T} is $T=4p^6+4p^4-2$. Its three pairs of roots are distinct and nonzero, and $\pi_pW_x=1$ at each ramification point, so the curve lies in the regular domain. The genus-one field correction is
\begin{equation}\label{eq:D3Phi}
 \Phi_1=-\frac{398}{2187}+\frac{884}{2187}p^2-\frac{500}{2187}p^{-2},
\end{equation}
and the reconstruction gives the primitive and the one-holed torus
\begin{equation}\label{eq:D3q}
 \Qhat_1^D=-\frac{8\,p^3N_D}{81\,T^5},\qquad \omega_{1,1}^D=-\frac{8\,N_D}{81\,T^4}\,\dd p,
\end{equation}
with
\begin{align}\label{eq:ND}
 N_D={}&176p^{22}+440p^{20}+100p^{18}-1616p^{16}-1172p^{14}+336p^{12}\nonumber\\
 &-1616p^{10}-1430p^8-245p^6-90p^4+24p^2-10 .
\end{align}
The same function follows from the corrected trace \eqref{eq:correctedtrace} of the string and from the velocity formula \eqref{eq:Qstar}. The scalar calculation, with $E_A'(k)=4k^3+2k-i$, gives
\begin{equation}\label{eq:A3q}
 \Qhat_1^A=-\frac{N_A}{81\,E_A'(k)^5},\qquad \omega_{1,1}^A=-\frac{N_A}{81\,E_A'(k)^4}\,\dd k,
\end{equation}
\begin{equation}\label{eq:NA}
 N_A=704k^{10}+640ik^9+880k^8+640ik^7+620k^6+104ik^5-238k^4-60ik^3-53k^2-72ik-17 .
\end{equation}

The two endpoint conditions of the pencil can be read off directly,
\begin{equation}
 \Qhat_1^D=-\frac5{162}\,p^3+O(p^5)\ \ (p\to0),\qquad \Qhat_1^D=-\frac{11}{648}\,p^{-5}+O(p^{-7})\ \ (p\to\infty).
\end{equation}
They also show how the quantum integration constants are fixed. Allowing order-$\h^2$ shifts of $(\mathfrak c_0,\mathfrak c_1,\gamma_0)$ changes the primitive by
\begin{equation}
 \delta\Qhat_1^D=\delta\mathfrak c_0\,\frac{p^3}{T}+\delta\mathfrak c_1\,\frac{3p^3(2p^2+1)}{2T}+\delta\gamma_0\,\frac{2p}{T} .
\end{equation}
At infinity these shifts produce the forbidden modes $p^{-1}$ and $p^{-3}$, and at zero the shift of $\gamma_0$ produces the forbidden linear term,
\begin{equation}\label{eq:D3modes}
 \begin{array}{c|ccc}
  & \delta\mathfrak c_0 & \delta\mathfrak c_1 & \delta\gamma_0\\ \hline
 p^{-1}\ \text{at}\ \infty & 0 & 3/4 & 0\\
 p^{-3}\ \text{at}\ \infty & 1/4 & -3/8 & 0\\
 p\ \text{at}\ 0 & 0 & 0 & -1
 \end{array}
 \qquad \det=\frac3{16} .
\end{equation}
The matrix is invertible, so the three constants are fixed uniquely, the two ordinary ones at infinity and the Pfaffian one at the origin. This is the mechanism of section~\ref{sec:strings} in its smallest form.
\section{Potts and tetracritical Ising in detail}\label{app:potts}

This appendix records the data behind section~\ref{sec:potts}. Everything is evaluated on the family \eqref{eq:pottsfamily} at $x=1$ unless stated otherwise.

\paragraph{Ordering and sources.}

The two operators \eqref{eq:pottsops} share their leading fields,
\begin{equation}
 (u^A_{0,0},u^A_{2,0},u^A_{4,0})=(u_{1,0},u_{2,0},u_{3,0})=\bigl(-2x^{3/5},\,9x^{2/5},\,-6x^{1/5}\bigr),
\end{equation}
but their differential powers produce different ordering terms,
\begin{equation}\label{eq:pottsordering}
 O_A=\frac32u^A_{4,0}{}''p^2+\frac14u^A_{2,0}{}''=\frac{36}{25}p^2-\frac{27}{50},\qquad
 O_D=\frac52u_{3,0}''p^3+\frac34u_{2,0}''p=\frac{12}5p^3-\frac{81}{50}p .
\end{equation}
With $W_x=-6(p^4-3p^2+1)/5$ and $W_{xx}=(24p^4-54p^2+12)/25$ at $x=1$, the numerators of the inverse symbol over powers of $W-\lambda$, including the momentum insertion for the pencil, are
\begin{align}
 f_2^A&=-O_A, & f_2^D&=-\frac{O_D}p-\frac{W_{xx}}{4p^2},\nonumber\\
 f_3^A&=\frac14\bigl(W_{pp}W_{xx}-W_{px}^2\bigr), & f_3^D&=f_3^A+\frac{W_x^2}{4p^2},\\
 f_4&=-\frac14\bigl(W_{pp}W_x^2-2W_{px}W_pW_x+W_{xx}W_p^2\bigr),\nonumber
\end{align}
where $f_4$ is common to the two theories. The residue rule \eqref{eq:residuerule} turns them into the genus-one sources.

\paragraph{Genus one.}

The homogeneous primitive $\Qhat_1=x^{-17/10}q(z)$ obeys
\begin{equation}
 -\frac{17}{10}q-\frac{3w}{5w'}q'=\mathcal S_1+\frac{\Phi w''-\Phi'w'}{(w')^3},
\end{equation}
with $\Phi=\phi_{1,0}+\phi_{1,2}z^2+\phi_{1,4}z^4$. With $Z=z^2$, $\mathcal C=Z^2-4Z+3$ and $q=N(Z)/(720z^5\mathcal C^5)$, clearing denominators gives
\begin{align}
 &(8Z^5-74Z^4+226Z^3-272Z^2+102Z-30)N\nonumber\\
 &\qquad-(2Z^6-20Z^5+72Z^4-112Z^3+70Z^2-12Z)N'\nonumber\\
 &\qquad-200Z^2\mathcal C^4\bigl[(\mathcal C+2Z\mathcal C')\Phi-2Z\mathcal C\Phi'\bigr]=7200z^7\mathcal C^7\mathcal S_1,
\end{align}
with primes now denoting $Z$-derivatives. Its fifteen coefficients determine the ten numerator coefficients and three fields of table~\ref{tab:pottsg1}. The constant equation alone gives $N(0)=-81$ for A and $-324$ for D. The numerators of \eqref{eq:pottsW11} are
\begin{align}\label{eq:pottsN}
 \mathcal N_A={}&10z^{18}-121z^{16}+601z^{14}-1562z^{12}+2317z^{10}\nonumber\\
 &-1872z^8+387z^6+594z^4-243z^2+81,\nonumber\\
 \mathcal N_D={}&4z^{18}-52z^{16}+280z^{14}-797z^{12}+1279z^{10}\nonumber\\
 &-1065z^8+174z^6+441z^4-297z^2+81 .
\end{align}

\paragraph{Through genus three.}

An independent route uses the two string generators,
\begin{align}
 S_A&=D^5+\frac12\Bigl\{\frac56u^A_4,D^3\Bigr\}\nonumber\\
 &\quad+\frac12\Bigl\{\frac56u^A_2-\frac5{72}(u^A_4)^2+\frac{25}{72}\h^2u^A_4{}'',D\Bigr\},\nonumber\\
 S_D&=D^5+\frac56u_3D^3+\frac56\h u_3'D^2\nonumber\\
 &\quad+\Bigl(\frac56u_2-\frac5{72}u_3^2+\frac{20}9\h^2u_3''\Bigr)D,
\end{align}
with $[S_A,Q_A]=-6\h$ and $DS_DD^{-1}L_D-L_DS_D=-6\h D$. At genus $g$ write the new fields as $\phi_{g,0}x^{(3-11g)/5}$, $\phi_{g,2}x^{(2-11g)/5}$ and $\phi_{g,4}x^{(1-11g)/5}$, and put $\mathsf m=11g-3$. With source components $\Delta_{g,k}$ obtained by keeping every lower-genus field and setting the new ones to zero, in both theories they solve
\begin{gather}
 -\mathsf m\phi_{g,0}-\phi_{g,2}-\phi_{g,4}+\Delta_{g,0}=0,\qquad 3\mathsf m\phi_{g,0}+(2\mathsf m+1)\phi_{g,2}+(3\mathsf m+3)\phi_{g,4}+\Delta_{g,2}=0,\nonumber\\
 -\mathsf m\phi_{g,0}-\mathsf m\phi_{g,2}-2\mathsf m\phi_{g,4}+\Delta_{g,4}=0,
\end{gather}
whose solution is
\begin{gather}
 \phi_{g,4}=\frac{\Delta_{g,0}+\Delta_{g,2}+2\Delta_{g,4}}{\mathsf m-2},\qquad \phi_{g,2}=\frac{\Delta_{g,4}-\Delta_{g,0}-(2\mathsf m-1)\phi_{g,4}}{\mathsf m-1},\nonumber\\
 \phi_{g,0}=\frac{\Delta_{g,0}-\phi_{g,2}-\phi_{g,4}}{\mathsf m} .
\end{gather}
Table~\ref{tab:pottsfields} lists the sources together with the fields. The Weyl-symbol reconstruction of section~\ref{sec:recipe} gives the same fields, with coefficient systems of sizes $30\times28$ and $45\times43$ at genera two and three, whose cleared residuals vanish identically.

\begin{table}[t]
\centering
\small
\begin{tabular}{@{}ccrrrrrr@{}}
\toprule
& $g$ & $\Delta_{g,0}$ & $\Delta_{g,2}$ & $\Delta_{g,4}$ & $\phi_{g,0}$ & $\phi_{g,2}$ & $\phi_{g,4}$\\
\midrule
A & 1 & $\frac{74}{25}$ & $\frac{213}{25}$ & $-\frac{106}{25}$ & $\frac{57}{100}$ & $-\frac{21}{10}$ & $\frac12$\\[3pt]
D & 1 & $\frac{88}5$ & $-\frac{288}{25}$ & $-\frac{16}{25}$ & $\frac{66}{25}$ & $-\frac{108}{25}$ & $\frac45$\\[3pt]
A & 2 & $\frac{5939057}{150000}$ & $\frac{327723}{50000}$ & $-\frac{560291}{50000}$ & $\frac{173249}{75000}$ & $-\frac{56923}{10000}$ & $\frac{2618}{1875}$\\[3pt]
D & 2 & $\frac{3159332}{9375}$ & $-\frac{1588692}{3125}$ & $\frac{352564}{3125}$ & $\frac{174088}{9375}$ & $-\frac{11884}{625}$ & $\frac{5984}{1875}$\\[3pt]
A & 3 & $\frac{1941485473}{2812500}$ & $-\frac{26007598}{234375}$ & $-\frac{32192723}{450000}$ & $\frac{824281321}{33750000}$ & $-\frac{144923317}{2500000}$ & $\frac{38951913}{2500000}$\\[3pt]
D & 3 & $\frac{6317841952}{703125}$ & $-\frac{713413792}{46875}$ & $\frac{105129568}{28125}$ & $\frac{647732752}{2109375}$ & $-\frac{63556064}{234375}$ & $\frac{3464736}{78125}$\\
\bottomrule
\end{tabular}
\caption{Sources and fields for tetracritical Ising (A) and Potts (D) on the family \eqref{eq:pottsfamily} through genus three. The top coefficients $\phi_{g,4}$ give the free energies of table~\ref{tab:pottsratios} through $u^A_{4,g}=6\partial_x^2F^A_g$ and $u_{3,g}=12\partial_x^2F^D_g$.}\label{tab:pottsfields}
\end{table}

\paragraph{The resolved origin in the one-holed torus.}

The two-variable potential \eqref{eq:pottsLG} has Hessian determinants $-36,-36,12,-36$ at its critical points $(0,3),(0,-3),(1,0),(3,0)$, so all four are Morse. To see how the collision of the four origin points contributes to $W_{1,1}$, consider first a model of the collision, with energy $W=p^2+p^{-2}$ and disk $y=p$, representatives $p=1$ and $p=i$, and local involutions $\sigma_1(p)=1/p$ and $\sigma_i(p)=-1/p$. The residue integrand of the mirror recursion, including the pullback of the second argument of the kernel, is
\begin{equation}
 \frac{\dfrac1{p_0-p}-\dfrac1{p_0+p}-\dfrac1{p_0-\sigma}+\dfrac1{p_0+\sigma}}{2(p-\sigma)(2p-2p^{-3})}\left[\frac1{(p-\sigma)^2}+\frac1{(p+\sigma)^2}\right]\sigma'(p),
\end{equation}
with $\sigma=\sigma_1(p)$ or $\sigma_i(p)$. Its poles have order three, and the two residues are
\begin{equation}
 \Res_{p=1}=-\frac{p_0^2(p_0^2+1)}{8(p_0^2-1)^4}\,\dd p_0,\qquad
 \Res_{p=i}=-\frac{p_0^2(p_0^2-1)}{8(p_0^2+1)^4}\,\dd p_0,
\end{equation}
with sum $-p_0^4(p_0^8+10p_0^4+5)\,\dd p_0/(4(p_0^4-1)^4)$. Each residue separately starts at $p_0^{-4}$ at large $p_0$, so the collision model has no inverse-square moment. For a general local energy $W=W_0+W_2p^2+W_4p^4+\cdots$ and disk $y=y_1p+y_3p^3+\cdots$, the same collision gives
\begin{equation}\label{eq:doubletresidue}
 -\frac1{4W_2y_1}\left[\frac1{p_0^4}-\Bigl(\frac{y_3}{y_1}+\frac{W_4}{W_2}\Bigr)\frac1{p_0^2}\right]\dd p_0 .
\end{equation}
The inverse-square term depends on the subleading coefficients, so the collision model alone is not enough. For Potts, $W_2=9$, $W_4=-6$, $y_1=5/6$ and $y_3=-5/6$, which gives \eqref{eq:pottsdoublet}. Adding the residues at the two nonzero representatives reproduces the Potts one-holed torus \eqref{eq:pottsW11}.

\paragraph{The two-boundary numerator.}

The numerator of \eqref{eq:pottsW12} is a symmetric polynomial,
\begin{equation}
 \mathcal N_{12}(Z_1,Z_2)=\sum_{i=0}^{14}c_{ii}Z_1^iZ_2^i+\sum_{0\leq i<j\leq14}c_{ij}\bigl(Z_1^iZ_2^j+Z_1^jZ_2^i\bigr).
\end{equation}
Each row below lists $(c_{ii},c_{i,i+1},\ldots,c_{i,14})$. The amplitude follows from two dispersionless insertions of the free energy on the normalized symmetric family, which is legitimate because insertion commutes with reconstruction.
{\footnotesize\begin{flushleft}
$i=0$\quad $(0,\allowbreak\ 0,\allowbreak\ 885735,\allowbreak\ -7085880,\allowbreak\ 25391070,\allowbreak\ -53800200,\allowbreak\ 74959425,\allowbreak\ -72258480,\allowbreak\ 49372740,\allowbreak\ -24086160,\allowbreak\ 8328825,\allowbreak\ -1992600,\allowbreak\ 313470,\allowbreak\ -29160,\allowbreak\ 1215)$\\[2pt]
$i=1$\quad $(531441,\allowbreak\ -9211644,\allowbreak\ 54915570,\allowbreak\ -174470112,\allowbreak\ 346256775,\allowbreak\ -463127868,\allowbreak\ 434271132,\allowbreak\ -290939040,\allowbreak\ 139879791,\allowbreak\ -47836980,\allowbreak\ 11346642,\allowbreak\ -1772928,\allowbreak\ 164025,\allowbreak\ -6804)$\\[2pt]
$i=2$\quad $(54305397,\allowbreak\ -209466486,\allowbreak\ 542745603,\allowbreak\ -970198398,\allowbreak\ 1226229273,\allowbreak\ -1114476732,\allowbreak\ 734162373,\allowbreak\ -350023032,\allowbreak\ 119304495,\allowbreak\ -28290222,\allowbreak\ 4427217,\allowbreak\ -410670,\allowbreak\ 17091)$\\[2pt]
$i=3$\quad $(547563564,\allowbreak\ -1083186108,\allowbreak\ 1664331786,\allowbreak\ -1958860098,\allowbreak\ 1731332151,\allowbreak\ -1131228450,\allowbreak\ 538536357,\allowbreak\ -183301704,\allowbreak\ 43271793,\allowbreak\ -6711498,\allowbreak\ 613899,\allowbreak\ -25056)$\\[2pt]
$i=4$\quad $(1649920914,\allowbreak\ -2225021724,\allowbreak\ 2668367448,\allowbreak\ -2586540762,\allowbreak\ 1875766113,\allowbreak\ -978324966,\allowbreak\ 357973506,\allowbreak\ -89197578,\allowbreak\ 14367051,\allowbreak\ -1345086,\allowbreak\ 55431)$\\[2pt]
$i=5$\quad $(3122497890,\allowbreak\ -4417736580,\allowbreak\ 4968507897,\allowbreak\ -3983380794,\allowbreak\ 2218858128,\allowbreak\ -849590100,\allowbreak\ 218939841,\allowbreak\ -36223074,\allowbreak\ 3469635,\allowbreak\ -145962)$\\[2pt]
$i=6$\quad $(7146856494,\allowbreak\ -8526087582,\allowbreak\ 7006759174,\allowbreak\ -3954475122,\allowbreak\ 1529378514,\allowbreak\ -398114048,\allowbreak\ 66625773,\allowbreak\ -6468654,\allowbreak\ 276571)$\\[2pt]
$i=7$\quad $(10298947932,\allowbreak\ -8481007560,\allowbreak\ 4794952260,\allowbreak\ -1861640964,\allowbreak\ 487752549,\allowbreak\ -82374702,\allowbreak\ 8092719,\allowbreak\ -351138)$\\[2pt]
$i=8$\quad $(6984533884,\allowbreak\ -3957675680,\allowbreak\ 1544078391,\allowbreak\ -407546774,\allowbreak\ 69500359,\allowbreak\ -6910086,\allowbreak\ 304150)$\\[2pt]
$i=9$\quad $(2252678857,\allowbreak\ -884851032,\allowbreak\ 235609071,\allowbreak\ -40607082,\allowbreak\ 4087392,\allowbreak\ -182462)$\\[2pt]
$i=10$\quad $(350544213,\allowbreak\ -94278228,\allowbreak\ 16433622,\allowbreak\ -1675008,\allowbreak\ 75810)$\\[2pt]
$i=11$\quad $(25636126,\allowbreak\ -4521704,\allowbreak\ 466698,\allowbreak\ -21406)$\\[2pt]
$i=12$\quad $(807308,\allowbreak\ -84372,\allowbreak\ 3920)$\\[2pt]
$i=13$\quad $(8928,\allowbreak\ -420)$\\[2pt]
$i=14$\quad $(20)$\\[2pt]
\end{flushleft}}
\section{More closed backgrounds}\label{app:homogeneous}

Potts is one member of a large family. In this appendix we collect the closed free energies of further D-series backgrounds, the height-model test, and the genus-three formula referred to in section~\ref{sec:highergenus}.

\paragraph{Ratios of free energies.}

Table~\ref{tab:ratios} compares the D and A free energies on common homogeneous families. The upper block uses Chebyshev endpoints $w=e_h$ and contains the unitary D-series models with $s=h\pm1$. The lower block uses $w=e_h+(h/s)e_{h-2s}$ with $3s=h\pm1$, which describes non-unitary D-series backgrounds. In every row the genus-one ratio is the count \eqref{eq:R1}, and in the upper block the genus-two ratio is \eqref{eq:R2}.

\begin{table}[t]
\centering
\begin{tabular}{@{}lccccc@{}}
\toprule
Model & $(h,s)$ & $c$ & $R_1$ & $R_2$ & $R_3$\\
\midrule
Potts & $(6,5)$ & $4/5$ & $4/5$ & $8/7$ & $98816/69433$\\
Tricritical Potts & $(6,7)$ & $6/7$ & $4/5$ & $8/7$ & $34448/24625$\\
$(A_6,D_5)$ & $(8,7)$ & $25/28$ & $5/7$ & $6/7$ & $452/541$\\
$(A_8,D_5)$ & $(8,9)$ & $11/12$ & $5/7$ & $6/7$ & $34372/41741$\\
$(A_8,D_6)$ & $(10,9)$ & $14/15$ & $2/3$ & $8/11$ & $246076/379907$\\
$(A_{10},D_6)$ & $(10,11)$ & $52/55$ & $2/3$ & $8/11$ & $1453504/2263253$\\
\midrule
$(A_2,D_5)$ & $(8,3)$ & $-21/4$ & $5/7$ & $14/13$ & $49748/25399$\\
$(A_2,D_6)$ & $(10,3)$ & $-44/5$ & $2/3$ & $16/17$ & $187328/111547$\\
$(A_4,D_8)$ & $(14,5)$ & $-208/35$ & $8/13$ & $16/23$ & $705568/892337$\\
$(A_4,D_9)$ & $(16,5)$ & $-323/40$ & $3/5$ & $2/3$ & $772948/1021379$\\
\bottomrule
\end{tabular}
\caption{Exact ratios $R_g=F^D_g/F^A_g$ of the closed free energies in the symmetric sector, from reconstruction on homogeneous families. The first six rows have $w=e_h$, and the last four have $w=e_h+(h/s)e_{h-2s}$.}\label{tab:ratios}
\end{table}

Potts and tricritical Potts share $e_6$ and their first two ratios, while their genus-three ratios differ. Genus three is the first order at which the partner degree and the motion of the background enter the ratio.

\paragraph{A height-model test.}

The height-model constructions of refs.~\cite{Kostov1992,HiguchiKostov1995} compute the unitary vacuum amplitudes of both theories as sums over graphs. The two genus-counting parameters may be normalized differently in the height model and in our construction, so we compare the combination
\begin{equation}\label{eq:heightinvariant}
 \mathcal I_g=\frac{F^D_g\,(F^D_0)^{g-1}}{F^A_g\,(F^A_0)^{g-1}}=R_gR_0^{g-1},\qquad R_0=\frac12,
\end{equation}
which is invariant under separate rescalings of the two. The genus-three enumeration applies the Higuchi--Kostov rules to the forty-two connected stable graphs, with fusion coefficients in $\mathbb Q(\sqrt3)$, intersection numbers of $\psi$ classes and graph automorphism factors. After removing a common height-model factor its coefficients are
\begin{equation}
 \bigl(\mathfrak h_3^{A_5},\mathfrak h_3^{D_4}\bigr)_{\rm dense}=\Bigl(\frac{1180361}{196830},\frac{209984}{98415}\Bigr),\qquad
 \bigl(\mathfrak h_3^{A_5},\mathfrak h_3^{D_4}\bigr)_{\rm dilute}=\Bigl(\frac{2339375}{59049},\frac{818140}{59049}\Bigr),
\end{equation}
and the resulting invariants
\begin{equation}
 \mathcal I_3^{\rm Potts}=\frac{24704}{69433},\qquad \mathcal I_3^{\rm tricritical\ Potts}=\frac{8612}{24625}
\end{equation}
are $R_3/4$, as \eqref{eq:heightinvariant} requires.

\paragraph{The genus-three formula.}

At genus three a sum of local terms no longer fits the data. One also needs the variation of one Hessian under a canonical deformation of another critical value. With the Hessians $w''(a)$, define
\begin{equation}\label{eq:intrinsic}
 \mathsf I_{1,a}=w''(a)\,\partial_a\log w''(a),\qquad \mathsf I_{2,a}=-\frac12\sum_{b\neq a}w''(b)\,\partial_b\log w''(a),
\end{equation}
where the canonical derivative $\partial_b$ raises the $b$th critical value by one and keeps all others fixed \cite{Dubrovin1996}. In D the inner sum runs over the $n$ canonical directions. On the regular D chart the canonical tangent is $\mathsf t_b(z)=2zw'(z)/(w''(b)(z^2-b^2))$, and the off-diagonal variation is
\begin{equation}
 \partial_b\log w''(a)=-\frac2{w''(b)}\left[\frac1{(a-b)^2}+\frac1{(a+b)^2}\right]\qquad(a\neq b),
\end{equation}
which again shows the paired structure. Put
\begin{equation}
 \alpha=\frac{2h}{h+2s},\qquad \mathcal J=\frac{h+2s}{2(h+s-1)} .
\end{equation}
Every reconstructed genus-three field we have computed agrees with
\begin{conjecture}[Genus three]\label{conj:g3}
On regular homogeneous backgrounds,
\begin{equation}\label{eq:g3}
 F_3^{\vecc}=x^{-2H/\mathfrak d}\,\mathcal J^4\sum_a\vartheta_a\Bigl[c_3\vartheta_a-(4\mathfrak m+3\mathfrak n)\mathsf I_{1,a}-(4\mathfrak m+6\mathfrak n)\mathsf I_{2,a}\Bigr],
\end{equation}
with the sum over all $2n$ momentum critical points and coefficient functions evaluated at $\alpha$,
\begin{align}
 c_3(\alpha)&=-\frac{(\alpha-1)(3\alpha+2)(31\alpha^3+189\alpha^2+196\alpha+4)}{1451520\,\alpha^2},\nonumber\\
 \mathfrak m(\alpha)&=-\frac{(\alpha+4)(3\alpha+2)(39\alpha^2-6\alpha-68)}{2903040\,\alpha},\\
 \mathfrak n(\alpha)&=\frac{(\alpha+4)(3\alpha+2)(5\alpha^2-4\alpha-8)}{967680\,\alpha}.\nonumber
\end{align}
\end{conjecture}
The genus-two law \eqref{eq:g2law} has the same structure with $\widetilde\Phi_2=\mathcal J^2c_2(\alpha)$ and $c_2(\alpha)=(\alpha-1)(3\alpha+2)/(2880\,\alpha)$. The coefficient $c_3$ is the Witten--Kontsevich genus-three free energy on a power-law solution, while $\mathfrak m$ and $\mathfrak n$ are fixed by the principal parts of a graph sum under a common shift of critical values. A polynomial remainder is not excluded by that argument, and its vanishing is part of what the conjecture asserts. In the symmetric sector the formula is applied to the combined resolved summands before the limit is taken.

\paragraph{Regular and nonsymmetric controls.}

The genus-three formula can also be tested away from conformal endpoints. Table~\ref{tab:controls} lists the exact vector top fields for three regular $D_4$ backgrounds and one $D_3$ background, on which the Pfaffian field is nonzero. They test the D form of \eqref{eq:g3}. The two scalar controls without reflection symmetry test the same intrinsic structure with the scalar deformation space. All values come from reconstruction in rational arithmetic, with critical-point sums evaluated as traces in the quotient algebra of the critical polynomial.

\begin{table}[t]
\centering
\small
\begin{tabular}{@{}lcc@{}}
\toprule
$w(z)$ & $(h,s)$ & coefficient of $u_{n-1,3}$\\
\midrule
$e_6(z)+z^{-2}$ & $(6,5)$ & $4456860012/76146875$\\
$z^6-6z^4+3z^2+4+9z^{-2}$ & $(6,5)$ & $1154912/546875$\\
$z^6+3z^4-4z^2+1+2z^{-2}$ & $(6,5)$ & $101100410352227308/7073934923671875$\\
$z^4-8z^2+12-4z^{-2}$ & $(4,3)$ & $2275641148/80838081$\\
\midrule
$z^6-5z^4+\tfrac12z^3-2z^2+z+3$ & $(6,5)$ & $-\frac{124584504217462026850673333333}{18092454444936408457628750000}$\\[3pt]
$z^6-4z^4-z^3+7z^2+2z-1$ & $(6,5)$ & $\frac{3995155602055877192989}{338522933668504375000}$\\
\bottomrule
\end{tabular}
\caption{Genus-three vector top fields on regular D backgrounds (upper block) and on two scalar backgrounds without reflection symmetry (lower block). The last column is the coefficient of the homogeneous power of $x$ in the top field $u_{n-1,3}$.}\label{tab:controls}
\end{table}
\section{Comparison with the worldsheet}\label{app:worldsheet}

Rodriguez, Usatyuk and Wang computed sphere four-point and torus one-point amplitudes of ADE minimal strings by direct numerical integration over the moduli space \cite{RodriguezUsatyukWang2025}. This appendix compares every entry of their A- and D-series torus tables that lies in our primary range, together with ten mixed D-series sphere amplitudes, sketches a proof of their conjectured torus formulas in that range, and adds predictions for the pure Pfaffian torus amplitude, conditional on conjecture~\ref{conj:pfaffian}. Their diagonal sphere four-point amplitudes are not compared here.

\subsection{Coupling frame and sphere amplitudes}

We work at the Liouville conformal point $W=e_h$, $\pi=e_s$ and $\rho_0=0$, with $3\leq s<2h$ and $\gcd(h,s)=1$. For the D series $h$ is even and $s$ is odd, while the A-series statements below also allow odd $h$. The hodograph equations \eqref{eq:hodograph} define solution coordinates $\gamma_k$ and $\gamma_0$, and the ordinary flows act on them as
\begin{equation}\label{eq:timedictionary}
 \partial_{t_m}\boldsymbol u=-\frac{\kappa m}{h}\,\partial_{\gamma_{h-m}}\boldsymbol u .
\end{equation}
A diagonal tachyon with odd labels $(r,s')$, $1\leq r\leq h-1$, $1\leq s'\leq s-1$, has index $m=|sr-hs'|$. In the primary range $0<m<h$ with $s\nmid m$ we identify its coupling with $\gamma_{h-m}$. The bound $s<2h$ puts the dressed identity $(1,1)$, of index $|s-h|$, in this range. The first Pfaffian primary has $m=h/2$ and couples to $\gamma_0$.

Genus-zero amplitudes are residues on the two-variable potential \eqref{eq:LG},
\begin{equation}\label{eq:threetensor}
 \langle f_1,f_2,f_3\rangle=\sum_{\rm crit\,\cF}\frac{f_1f_2f_3}{\det\operatorname{Hess}\cF},
\end{equation}
which for ordinary tangents at $\rho=0$ becomes $\langle f_1,f_2,f_3\rangle=-\tfrac12[t^{-1}]\,f_1f_2f_3/(tA'(t))$. With $U_a$ the solution tangent of $\gamma_a$, $E_b$ the flat insertion vector of the flat coordinate $\mathsf v_b=[p^{-1}]W^{b/h}$, and $\pi_1=\pi_p(0)$, we take
\begin{equation}\label{eq:bulkprescription}
 C_{abc}=\langle U_a,E_b,E_c\rangle,\qquad C_{abcd}=\partial_{\gamma_d}C_{abc},\qquad
 C_{\mathrm{NN}b}=\frac{2E_b(0)}{\pi_1}.
\end{equation}
In the mixed amplitude only the resolved origin pair contributes, each point supplying $E_b(0)/\pi_1$. The normalized amplitudes are
\begin{align}\label{eq:amplitudeframe}
 T_{abc}&=\frac{\beta_a\beta_b\beta_c}{C_{III}}C_{abc}, & T_{abcd}&=\frac{\Lambda^{-1}}{C_{III}}\beta_a\beta_b\beta_c\beta_dC_{abcd},\\
 T_a^{\rm torus}&=K\beta_aF_{1,a}, & T_{ab}^{\rm torus}&=K\Lambda^{-1}\beta_a\beta_bF_{1,ab},\nonumber
\end{align}
where $F_{1,a}$ and $F_{1,ab}$ are derivatives of $F^D_1$ in the coupling frame and $I$ labels the identity. Fusion-allowed three-point amplitudes fix the operator factors $\beta_a$, one identity four-point value fixes $\Lambda$, and the identity torus value fixes $K$.

For $s=3$ the calibration conditions are $T_{III}=1$, $T_{IIII}=4s-2h$ and $T_I^{\rm torus}=h+2$. Table~\ref{tab:calibration} lists the resulting data, and table~\ref{tab:sphere} gives the mixed amplitudes with two Pfaffian and two diagonal legs, obtained by differentiating $C_{\mathrm{NN}b}$ along the second diagonal coupling. Ten of the twelve entries are in the worldsheet table, and all ten agree.

\begin{table}[t]
\centering
\begin{tabular}{@{}ccccc@{}}
\toprule
$(h,s)$ & $C_{III}$ & $C_{IIII}$ & $(\beta_1,\beta_3,\beta_5)$ & $\beta_{\mathrm N}^2$\\
\midrule
$(8,3)$ & $64/3$ & $128/9$ & $(1,-1,-1)$ & $-4$\\
$(10,3)$ & $100/3$ & $400/9$ & $(1,1,-1)$ & $5$\\
$(14,3)$ & $196/3$ & $1568/9$ & $(1,1,-1)$ & $7$\\
$(16,3)$ & $256/3$ & $2560/9$ & $(1,1,1)$ & $-8$\\
\bottomrule
\end{tabular}
\caption{Sphere calibration data at $s=3$. A negative $\beta_{\mathrm N}^2$ means an imaginary rescaling of the Pfaffian operator in that frame.}\label{tab:calibration}
\end{table}

\begin{table}[t]
\centering
\begin{tabular}{@{}ccccccc@{}}
\toprule
$(h,s)$ & $(1,1)$ & $(1,3)$ & $(1,5)$ & $(3,3)$ & $(3,5)$ & $(5,5)$\\
\midrule
$(8,3)$ & $-2$ & $-2$ & $-4$ & $-2$ & $-2$ & $-8^{\dagger}$\\
$(10,3)$ & $-4$ & $-2$ & $-2$ & $-2$ & $-2^{\dagger}$ & $-2$\\
\bottomrule
\end{tabular}
\caption{Normalized mixed amplitudes $T_{\mathrm{NN}D_rD_{r'}}$ with $D_r=O_{r,1}$, indexed by $(r,r')$. The unmarked entries agree with ref.~\cite[Table~6]{RodriguezUsatyukWang2025}. Daggers mark predictions in the same frame. The remaining worldsheet entry at $(10,3)$ involves $O_{7,1}$, whose index exceeds $h$.}\label{tab:sphere}
\end{table}

\subsection{Torus one-point amplitudes}

Write $\epsilon={\rm sgn}(sr-hs')$, $L=(r-1)/2$, and let $\mathfrak r_\nu\in\{0,\ldots,h-1\}$ be the residue of $\nu s$ modulo $h$. We use the sphere convention in which fusion-allowed diagonal three-point amplitudes are one, and fix the remaining constant per model by the identity value $T^D_{1,1}=(s-1)(h+2)/2$.

\begin{proposition}[Torus one-point amplitudes]\label{prop:torusonepoint}
In the primary range, the derivatives of the genus-one free energies \eqref{eq:Fone} and \eqref{eq:FoneA} give
\begin{align}\label{eq:TDclosed}
 T^D_{r,s'}&=\frac3h\Bigl[\frac{(h-r)(h^2+s^2+2-m^2)}{12}-\frac{r(h^2-1)}6+(-1)^L\frac{h^2+s^2-m^2}4\nonumber\\
 &\hspace{6em}-\frac h2+2\!\!\sum_{\substack{|\nu|\leq L\\ \nu\ {\rm odd}}}\!\!\mathfrak r_\nu(h-\mathfrak r_\nu)\Bigr],\\
 T^A_{r,s'}&=\frac3h\Bigl[\frac{(h-r)(h^2+s^2+2-m^2)}6-\frac{r(h^2-1)}3+2\sum_{|\nu|\leq L}\mathfrak r_\nu(h-\mathfrak r_\nu)\Bigr],\label{eq:TAclosed}
\end{align}
and these equal the formulas conjectured in ref.~\cite{RodriguezUsatyukWang2025},
\begin{equation}\label{eq:RUW}
 T^A_{r,s'}=\sum_{\bar r\in \mathfrak S_h(r)}\sum_{\bar s\in \mathfrak S_s(s')}\mathcal K(\bar r,\bar s),\qquad
 T^D_{r,s'}=\sum_{\substack{\bar r\in \mathfrak S_h(r)\\ \bar r\ {\rm odd}}}\sum_{\bar s\in \mathfrak S_s(s')}\mathcal K(\bar r,\bar s)+(-1)^L\sum_{\bar s\in \mathfrak S_s(s')}\mathcal K(h/2,\bar s),
\end{equation}
with $\mathcal K(\bar r,\bar s)=2hs-rs-s'h-6|\bar rs-\bar sh|$ and $\mathfrak S_h(r)=\{(r+1)/2,\ldots,h-(r+1)/2\}$.
\end{proposition}

The proof below is a sketch. The D formula uses propositions~\ref{thm:symmetriclimit} and~\ref{prop:quantum-tau}, whose proofs are also sketched, and the A formula in addition uses the scalar counterpart of proposition~\ref{prop:quantum-tau}. The pair $(\bar r,\bar s)$ labels an intermediate diagonal field, and $\mathfrak S_h(r)$ collects the labels whose self-fusion contains $r$. The scalar formula holds for every coprime pair with $s<2h$, and the D formula in addition requires $h$ even. The A-series statements use the first derivatives of $F^A_1$, which rest on the scalar counterpart of proposition~\ref{prop:quantum-tau} discussed at the end of appendix~\ref{app:genusone}.

\paragraph{Sketch of the proof.} Let $a_j=2\cos\theta_j$ with $\theta_j=\pi j/h$ be the critical points, write primes for momentum derivatives, and let $\delta W$ and $\delta\pi$ be the variations of $W$ and $\pi$ along $(m/h)\partial_{\gamma_{h-m}}$. The flow and the string equation give $\delta W(a_j)=-e_m'(a_j)/\pi'(a_j)$ with $\deg\delta W\leq h-2$, and the decay of the fixed-energy variation of $\pi$ gives the polynomial identity $W'\,\delta\pi-\pi'\,\delta W=e_m'$. The steps are then as follows.
\begin{enumerate}
 \item \emph{Normalization.} In canonical coordinates the sphere three-point function reduces, up to the product of the signs $\epsilon_i$, to the level-$(h-2)$ Verlinde sum $\sum_j\prod_i\sin(r_is\theta_j)/\sin(s\theta_j)$. With an identity leg this is $\sum_j\sin^2(rs\theta_j)=h/2$ for every $r$, so every operator has the same normalization magnitude, and the relative signs are $\epsilon\,{\rm sgn}(s-h)$. The identity direction is the scaling $W_\ell(p)=e^{h\ell}W(e^{-\ell}p)$, along which every velocity has the same logarithmic derivative. The identity torus value then fixes the overall scale to $K\beta_{r,s'}=48s\epsilon$.
 \item \emph{The velocity sum as a residue.} Let $\sigma_j$ be the local deck transformation at $a_j$ and define
 \begin{equation}
  \varpi_j=\Res_{p=a_j}\frac{e_m(p)-e_m(\sigma_jp)}{\pi(p)-\pi(\sigma_jp)}\,\frac{\dd p\,\dd(\sigma_jp)}{2(p-\sigma_jp)^2\,\dd W(p)} .
 \end{equation}
 A local expansion shows that $24\varpi_j$ differs from the variation of $\log v_j$ by half the residue at $a_j$ of $(\delta W''/W'-\delta W\,W'''/W'^2)\,\dd p$. That differential has no other poles and decays as $p^{-3}$, so its residues add up to zero, and $\sum_j\delta\log v_j=24\sum_j\varpi_j$.
 \item \emph{Trigonometric sums.} In the coordinate $p=2\cos(\theta_j+\mathsf u)$ the deck transformation is $\mathsf u\to-\mathsf u$, and
 \begin{equation}
  \varpi_j=\frac{m\sin m\theta_j}{16sh^2(-1)^j\sin s\theta_j}\Bigl[\frac{h^2+s^2+2-m^2}6-\csc^2\theta_j\Bigr].
 \end{equation}
 With $(-1)^j\sin m\theta_j=\epsilon\sin rs\theta_j$ and $\sin rs\theta/\sin s\theta=\sum_{|\nu|\leq L}\cos2\nu s\theta$, the sum over $j$ follows from $\sum_j\cos2k\theta_j=-1$ and $\sum_j\cos(2k\theta_j)\csc^2\theta_j=(h^2-1)/3-2k(h-k)$ for $0<k<h$. This gives \eqref{eq:TAclosed}. For odd $h$ the fixed-energy variation of $\pi$ may contain a constant at infinity, which changes $\delta\pi$ by a constant and leaves every $\delta\log v_j$ unchanged, so the same steps apply.
 \item \emph{The origin.} The point $p=0$ stays critical, and its velocity varies by $\delta\log v_{\rm fix}=-\delta\pi'(0)/\pi'(0)$. Comparing the coefficients of $p^0$ and $p^2$ in $W'\delta\pi=\pi'\delta W+e_m'$, and using $e_k'''(0)/e_k'(0)=(1-k^2)/4$ for odd $k$, gives
 \begin{equation}\label{eq:originvelocity}
  \delta\log v_{\rm fix}=-\frac1{2W''(0)}\Bigl[\delta W''(0)+\frac{s^2-m^2}4\,\frac{e_m'(0)}{\pi'(0)}\Bigr].
 \end{equation}
 The polynomial $\delta W$ is the interpolant of degree at most $h-2$ of its nodal values, and with $e_k(a_j)=2\cos k\theta_j$ these read $\delta W(a_j)=-(m\epsilon/s)\sum_{|\nu|\leq L}\cos\bigl((h+2\nu s)\theta_j\bigr)$. Let $d_\nu$ be the distance of $\nu s$ from $h\mathbb Z$. At every node the frequency $h+2\nu s$ has the same cosine as $h-2d_\nu$, and only $\nu=0$ gives the frequency $h$ itself, since $s$ is coprime to $h$ and $L<h/2$. The corresponding polynomial $e_h/2$ has degree $h$. Subtracting $pW'/(2h)$ removes its leading power and preserves every nodal value, because $W'$ vanishes at the nodes. The interpolant is therefore
 \begin{equation}\label{eq:origininterpolant}
  \delta W(p)=-\frac{m\epsilon}{2s}\Bigl[\sum_{|\nu|\leq L}e_{h-2d_\nu}(p)-\frac ph\,W'(p)\Bigr].
 \end{equation}
 With $e_{2b}''(0)=-2(-1)^bb^2$ it gives $\delta W''(0)=(m\epsilon/s)(-1)^{h/2}\bigl[\sum_{|\nu|\leq L}(-1)^\nu(h/2-d_\nu)^2-h/2\bigr]$, and the $-h/2$ comes from the degree correction. Insert this into \eqref{eq:originvelocity}, use $e_m'(0)/\pi'(0)=(m\epsilon/s)(-1)^{h/2+L}$ and $(h/2-d_\nu)^2=h^2/4-\mathfrak r_\nu(h-\mathfrak r_\nu)$, and multiply by $(h/m)K\beta_{r,s'}/16$. This gives the origin term
 \begin{equation}\label{eq:originonepoint}
  T^D_{r,s'}-\frac12T^A_{r,s'}=\frac3h\Bigl[(-1)^L\frac{h^2+s^2-m^2}4-\frac h2-\sum_{|\nu|\leq L}(-1)^\nu\mathfrak r_\nu(h-\mathfrak r_\nu)\Bigr],
 \end{equation}
 and adding half of \eqref{eq:TAclosed} gives \eqref{eq:TDclosed}.
 \item \emph{The worldsheet sums.} In the primary range, splitting the sum over $\bar s$ at $\lfloor\bar rs/h\rfloor$ gives $\sum_{\bar s}|\bar rs-\bar sh|=\tfrac h4[(s-s')^2-1]+s^2k^2/h+\mathfrak r(h-\mathfrak r)/h$, where $\bar r=h/2+k$ and $\mathfrak r$ is the residue of $\bar rs$ modulo $h$. Multiplication by $s$ permutes the residues modulo $h$, so the residue terms reorganize into the sums over $\nu$ in \eqref{eq:TDclosed} and \eqref{eq:TAclosed}, and what remains is a polynomial identity in $(h,s,r,s')$.
\end{enumerate}
Each step was also checked numerically. The residue form of step two holds to twenty-six digits for all $895$ primary labels with $h\leq32$ and $s\leq21$, and a symbolic check confirms the local identity for arbitrary Taylor data. The closed formulas equal the worldsheet sums \eqref{eq:RUW} in exact arithmetic for $6093$ D labels with $h\leq60$ and $s<40$, and for $10988$ A labels including odd $h$.

In this convention the identity rows are inputs, so the ratio $T^D_{1,1}/T^A_{1,1}=(h+2)/(2(h-1))$ is not a test. The corresponding prediction is the ratio $R_1=F_1^D/F_1^A$ of the vacuum free energies, which is fixed by the count \eqref{eq:R1} and is confirmed by independent reconstruction in tables~\ref{tab:pottsratios} and~\ref{tab:ratios}.

Tables~\ref{tab:torusD} and~\ref{tab:torusA} list every entry of the worldsheet tables of ref.~\cite{RodriguezUsatyukWang2025} that lies in the primary range. In each case our value and the worldsheet value coincide. Of the $176$ entries, $41$ are the identity rows, one for each model, and these fix the normalization. The remaining $135$ entries, $52$ for D and $83$ for A, are independent tests. The worldsheet tables also contain labels with $m>h$, which we do not attempt. They couple to gravitational descendants in the hierarchy frame, and the worldsheet integrals receive residues from poles crossing the Liouville contour once $m$ exceeds $h+s$.

For $s=3$ the labels of the worldsheet tables all have $s'=1$, and \eqref{eq:TAclosed} reduces to $T^A_{r,1}=1+2hr-3r^2$. Spodyneiko found the same dependence on $h$ and $r$ earlier for the $(3,p)$ minimal strings, from the matrix model after the resonance transformation to the Liouville frame \cite{Spodyneiko:2014lla,Spodyneiko:2015xpa}. In his notation $p=h$ and the label is $k=r-1$, and the result is stated for even $p$.

\begin{table}[p]
\centering
\small
\begin{tabular}{@{}l>{\raggedright\arraybackslash}p{0.8\linewidth}@{}}
\toprule
$(h,s)$ & primary-range torus one-point amplitudes $T^D_{r,s'}$\\
\midrule
$(8,3)$ & $T_{1,1}=10$, $T_{3,1}=14$, $T_{5,1}=6$\\
$(10,3)$ & $T_{1,1}=12$, $T_{3,1}=20$, $T_{5,1}=16$\\
$(14,3)$ & $T_{1,1}=16$, $T_{3,1}=32$, $T_{5,1}=36$, $T_{7,1}=28$, $T_{9,1}=8$\\
$(16,3)$ & $T_{1,1}=18$, $T_{3,1}=38$, $T_{5,1}=46$, $T_{7,1}=42$, $T_{9,1}=26$\\
$(20,3)$ & $T_{1,1}=22$, $T_{3,1}=50$, $T_{5,1}=66$, $T_{7,1}=70$, $T_{9,1}=62$, $T_{11,1}=42$, $T_{13,1}=10$\\
$(22,3)$ & $T_{1,1}=24$, $T_{3,1}=56$, $T_{5,1}=76$, $T_{7,1}=84$, $T_{9,1}=80$, $T_{11,1}=64$, $T_{13,1}=36$\\
$(6,5)$ & $T_{1,1}=16$, $T_{3,3}=0$\\
$(8,5)$ & $T_{1,1}=20$, $T_{3,1}=12$, $T_{5,3}=18$\\
$(12,5)$ & $T_{1,1}=28$, $T_{3,1}=36$, $T_{5,3}=14$, $T_{7,3}=26$, $T_{9,3}=18$\\
$(14,5)$ & $T_{1,1}=32$, $T_{3,1}=48$, $T_{5,1}=24$, $T_{7,3}=12$, $T_{9,3}=16$, $T_{11,3}=0$\\
$(16,5)$ & $T_{1,1}=36$, $T_{3,1}=60$, $T_{5,1}=44$, $T_{7,3}=6$, $T_{9,3}=22$, $T_{11,3}=18$\\
$(6,7)$ & $T_{1,1}=24$, $T_{3,3}=0$, $T_{5,5}=4$\\
$(8,7)$ & $T_{1,1}=30$, $T_{3,3}=4$, $T_{5,5}=6$\\
$(10,7)$ & $T_{1,1}=36$, $T_{3,3}=16$, $T_{5,3}=24$, $T_{7,5}=0$\\
$(12,7)$ & $T_{1,1}=42$, $T_{3,1}=30$, $T_{5,3}=44$, $T_{7,5}=22$, $T_{9,5}=30$\\
\bottomrule
\end{tabular}
\caption{D-series torus one-point amplitudes in the primary range for all models of ref.~\cite[Table~3]{RodriguezUsatyukWang2025}. Every entry equals both the worldsheet value and \eqref{eq:TDclosed}. The $T_{1,1}$ entries fix the normalization of each model, and the other $52$ entries are tests.}\label{tab:torusD}
\end{table}

\begin{table}[p]
\centering
\small
\begin{tabular}{@{}l>{\raggedright\arraybackslash}p{0.8\linewidth}@{}}
\toprule
$(h,s)$ & primary-range torus one-point amplitudes $T^A_{r,s'}$\\
\midrule
$(4,3)$ & $T_{1,1}=6$\\
$(5,3)$ & $T_{1,1}=8$, $T_{3,1}=4$\\
$(7,3)$ & $T_{1,1}=12$, $T_{3,1}=16$\\
$(8,3)$ & $T_{1,1}=14$, $T_{3,1}=22$, $T_{5,1}=6$\\
$(10,3)$ & $T_{1,1}=18$, $T_{3,1}=34$, $T_{5,1}=26$\\
$(11,3)$ & $T_{1,1}=20$, $T_{3,1}=40$, $T_{5,1}=36$, $T_{7,1}=8$\\
$(13,3)$ & $T_{1,1}=24$, $T_{3,1}=52$, $T_{5,1}=56$, $T_{7,1}=36$\\
$(14,3)$ & $T_{1,1}=26$, $T_{3,1}=58$, $T_{5,1}=66$, $T_{7,1}=50$, $T_{9,1}=10$\\
$(16,3)$ & $T_{1,1}=30$, $T_{3,1}=70$, $T_{5,1}=86$, $T_{7,1}=78$, $T_{9,1}=46$\\
$(17,3)$ & $T_{1,1}=32$, $T_{3,1}=76$, $T_{5,1}=96$, $T_{7,1}=92$, $T_{9,1}=64$, $T_{11,1}=12$\\
$(19,3)$ & $T_{1,1}=36$, $T_{3,1}=88$, $T_{5,1}=116$, $T_{7,1}=120$, $T_{9,1}=100$, $T_{11,1}=56$\\
$(20,3)$ & $T_{1,1}=38$, $T_{3,1}=94$, $T_{5,1}=126$, $T_{7,1}=134$, $T_{9,1}=118$, $T_{11,1}=78$, $T_{13,1}=14$\\
$(5,4)$ & $T_{1,1}=12$, $T_{3,3}=2$\\
$(7,4)$ & $T_{1,1}=18$, $T_{3,1}=12$, $T_{5,3}=6$\\
$(9,4)$ & $T_{1,1}=24$, $T_{3,1}=30$, $T_{5,3}=4$, $T_{7,3}=10$\\
$(11,4)$ & $T_{1,1}=30$, $T_{3,1}=48$, $T_{5,1}=18$, $T_{7,3}=12$, $T_{9,3}=14$\\
$(13,4)$ & $T_{1,1}=36$, $T_{3,1}=66$, $T_{5,1}=48$, $T_{7,3}=6$, $T_{9,3}=20$, $T_{11,3}=18$\\
$(15,4)$ & $T_{1,1}=42$, $T_{3,1}=84$, $T_{5,1}=78$, $T_{7,1}=24$, $T_{9,3}=18$, $T_{11,3}=28$, $T_{13,3}=22$\\
$(17,4)$ & $T_{1,1}=48$, $T_{3,1}=102$, $T_{5,1}=108$, $T_{7,1}=66$, $T_{9,3}=8$, $T_{11,3}=30$, $T_{13,3}=36$, $T_{15,3}=26$\\
$(6,5)$ & $T_{1,1}=20$, $T_{3,3}=6$\\
$(7,5)$ & $T_{1,1}=24$, $T_{3,3}=8$, $T_{5,3}=12$\\
$(8,5)$ & $T_{1,1}=28$, $T_{3,1}=12$, $T_{5,3}=18$\\
$(9,5)$ & $T_{1,1}=32$, $T_{3,1}=24$, $T_{5,3}=16$, $T_{7,3}=4$\\
$(11,5)$ & $T_{1,1}=40$, $T_{3,1}=48$, $T_{5,3}=12$, $T_{7,3}=24$\\
$(12,5)$ & $T_{1,1}=44$, $T_{3,1}=60$, $T_{5,3}=10$, $T_{7,3}=34$, $T_{9,3}=18$\\
$(13,5)$ & $T_{1,1}=48$, $T_{3,1}=72$, $T_{5,1}=16$, $T_{7,3}=36$, $T_{9,3}=32$\\
\bottomrule
\end{tabular}
\caption{A-series torus one-point amplitudes in the primary range for all models of ref.~\cite[Tables~1 and 2]{RodriguezUsatyukWang2025}, with $(h,s)=(p,p')$. Every entry equals both the worldsheet value and \eqref{eq:TAclosed}. The $T_{1,1}$ entries fix the normalization of each model, and the other $83$ entries are tests.}\label{tab:torusA}
\end{table}

\subsection{The pure Pfaffian torus amplitude}

The Pfaffian coupling is $\gamma_0$, and the first Pfaffian flow moves it at the rate \eqref{eq:origintangent}. The ordinary insertions do not reach this direction. Proposition~\ref{prop:quantum-tau} fixes $F^D_1$ only up to a function $\mathsf a(\gamma_0)$ that is constant along the ordinary times, and $\mathsf a''(\gamma_0)$ contributes directly to the pure Pfaffian amplitude. To go further we need a statement about the Pfaffian direction itself.

\begin{conjecture}[Pfaffian Hessian]\label{conj:pfaffian}
On normalized ordinary strings in the regular domain, with the ordinary constants held fixed,
\begin{equation}\label{eq:pfaffiancurvature}
 \partial_{\gamma_0}^2F^D_1=\partial_{\gamma_0}^2F_*=\frac{u_{1,1}+\partial_x^2u_{2,0}}{\kappa^2},
\end{equation}
and the identity continues to the symmetric point of proposition~\ref{thm:symmetriclimit}.
\end{conjecture}

The last expression is what the local two-point structure of the D tau function in the Pfaffian time gives \cite{LiuWuZhang2009}, provided the first Pfaffian flow translates $\gamma_0$ at the rate \eqref{eq:origintangent} while fixing the ordinary constants. The middle equality would then follow by comparing the genus-one field correction with the velocity sum, and with it $\mathsf a''(\gamma_0)=0$. We do not prove these steps here, and the amplitudes below are conditional on conjecture~\ref{conj:pfaffian}. At the conformal point the velocity product is the determinant of multiplication by the spatial tangent in the Jacobi algebra of \eqref{eq:LG}, and its second variation in $\gamma_0$ can be computed in closed form. For the strings $S=B_s$ with $s=3,5,7$, conjecture~\ref{conj:pfaffian} then gives the normalized two-point amplitudes
\begin{equation}\label{eq:pfaffianamplitudes}
 T_{\mathrm{NN}}^{(h,3)}=(-1)^{\lfloor h/6\rfloor}\frac{h^2+44}3,\qquad
 T_{\mathrm{NN}}^{(h,5)}=\frac{(-1)^{\lfloor h/10\rfloor}}5\begin{cases}4(h^2+146),& h/2\equiv\pm1,\\ 3(h^2+124),& h/2\equiv\pm2,\end{cases}
\end{equation}
\begin{equation}
 T_{\mathrm{NN}}^{(h,7)}=\frac{(-1)^{\lfloor h/14\rfloor}}7\begin{cases}11h^2+3148,& h/2\equiv\pm1,\\ 5h^2+1684,& h/2\equiv\pm2,\\ 6h^2+1464,& h/2\equiv\pm3,\end{cases}
\end{equation}
where the cases refer to the residue of $h/2$ modulo $s$. The mixed entries $T_{\mathrm Na}$ vanish by reflection in $\gamma_0$. Table~\ref{tab:pfaffian} lists the first values, including Potts and tricritical Potts. None of these amplitudes appears in the worldsheet tables, so they are predictions, conditional on conjecture~\ref{conj:pfaffian}. Direct differentiation of the full hodograph and flat-coordinate Jacobians reproduces the closed forms for $\partial_{\gamma_0}^2F_*$ in exact arithmetic for all admissible even $h\leq64$.

\begin{table}[t]
\centering
\begin{tabular}{@{}cccccc@{}}
\toprule
$s=3$ & $(8,3)$ & $(10,3)$ & $(14,3)$ & $(16,3)$ & $(20,3)$\\
$T_{\mathrm{NN}}$ & $-36$ & $-48$ & $80$ & $100$ & $-148$\\
\midrule
$s=5$ & $(6,5)$ Potts & $(8,5)$ & $(12,5)$ & $(14,5)$ & \\
$T_{\mathrm{NN}}$ & $96$ & $168$ & $-232$ & $-192$ & \\
\midrule
$s=7$ & $(6,7)$ tricritical Potts & $(8,7)$ & & & \\
$T_{\mathrm{NN}}$ & $240$ & $264$ & & & \\
\bottomrule
\end{tabular}
\caption{Pure Pfaffian torus two-point amplitudes predicted by conjecture~\ref{conj:pfaffian}, through \eqref{eq:pfaffiancurvature}, in the normalization \eqref{eq:amplitudeframe}. In every row the identity calibration uses $T_{IIII}=4\min(h,s)-2\max(h,s)$ and $T_I^{\rm torus}=(s-1)(h+2)/2$.}\label{tab:pfaffian}
\end{table}

\clearpage
\bibliographystyle{JHEP}
\bibliography{refs}
\end{document}